\documentclass[11pt]{article}
\pdfoutput=1
\usepackage[utf8x]{inputenc}
\usepackage{amssymb,amsmath,mathrsfs,enumerate}
\usepackage{graphicx,rotate,multicol, multirow}
\usepackage{cite}
\usepackage{braket}
\usepackage[normalem]{ulem}
\usepackage{wrapfig}
\usepackage[textfont=it,labelfont=bf]{caption}
\usepackage{bm}
\usepackage[colorlinks=true,
linkcolor=red,
urlcolor=blue,
citecolor=blue]{hyperref}
\usepackage{lineno}
\usepackage{color}
\usepackage{bbold}
\long\def\rpl#1!!#2!!{\textcolor{red}{#1} \textcolor{blue}{#2}}

\def\bar{\overline}
\def\tilde{\widetilde}

\makeatother

\newcommand{\bea}{\begin{eqnarray}}
\newcommand{\eea}{\end{eqnarray}}

\newcommand{\be}{\begin{equation}}
\newcommand{\ee}{\end{equation}}
\def\beq{\begin{equation}}
\def\eeq{\end{equation}}
\newcommand{\ba}{\begin{eqnarray}}
\newcommand{\ea}{\end{eqnarray}}
\def\ifmath#1{\relax\ifmmode #1\else $#1$\fi}

\newcommand{\g}{\,\mbox{GeV}}

\newcommand{\Z}[1]{\ensuremath{\mathbbm{Z}_{#1}}} 

\def\vb#1{\vbox to #1 pt{}}

\renewcommand{\Re}{\textrm{Re}}
\renewcommand{\Im}{\textrm{Im}}

\usepackage{physics}
\newcommand{\one}{\mathbb{1}}

\evensidemargin=\oddsidemargin
\allowdisplaybreaks

\title{\Large\bf 
	Large pseudoscalar Yukawa couplings in the complex 3HDM
}

\author{
	\sf 
	Rafael Boto$^{a,}$\footnote{rafael.boto@tecnico.ulisboa.pt},
	Luis Lourenco$^{a,}$\footnote{luis.lourenco.1510@tecnico.ulisboa.pt},
	Jorge C. Romão$^{a,}$\footnote{jorge.romao@tecnico.ulisboa.pt},
	Joao P. Silva$^{a,}$\footnote{jpsilva@cftp.ist.utl.pt}
	\\[3mm]
	\small\em
	$^a$ Centro de F\'isica Te\'orica de Part\'iculas-CFTP and Departamento de
	F\'isica,  Instituto Superior T\'ecnico,\\  \small\em
	Universidade de Lisboa, Av
	Rovisco Pais, 1, P-1049-001 Lisboa, Portugal \\ 
}

\date{}

\begin{document}
	
	
\maketitle
\renewcommand*{\thefootnote}{\arabic{footnote}}
\setcounter{footnote}{0}
	
\begin{abstract}
As LHC's Run3 unfolds, we peer deeper into the couplings of the 125GeV Higgs ($h_{125}$) and
also test for extra scalars.
Even before such extra scalars are found,
one can probe them by using the fact that theories with multiple
Higgs doublets allow for substantial
deviations of the couplings of $h_{125}$ from their SM values.
Indeed, the observed couplings already place stringent limits on such theories.
We investigate the curious possibility that $h_{125}$ couples to the top quark
mostly as a scalar, while it couples to the bottom quark mostly as a pseudoscalar.
This possibility was allowed by 2017 data for the so-called C2HDM;
a two Higgs doublet model with a single source of explicit CP violation.
It was shown recently that this possibility disappears
when using the full experimental data of 2024.
Here we discuss a three Higgs doublet with explicit CP violation (C3HDM),
and show that the curious CP-even/CP-odd $t$/$b$ possibility is partly resuscitated,
prompting further experimental exploration of this prospect.
\end{abstract}
	
	\maketitle
	
\section{Introduction}
\label{s:intro}
 
After the discovery at LHC by ATLAS \cite{ATLAS:2012yve} and
CMS \cite{CMS:2012qbp} of the 125GeV scalar ($h_{125}$),
we have entered the epoch of the precise determination of its properties.
And, given that one such scalar exists, one is also eager to learn
whether additional fundamental scalars exist in Nature.
These two ideas cross-fertilize: precise determination of $h_{125}$
couplings constrain the properties of theories with extra scalars;
theories with extra scalars point out avenues for exploration
of peculiar $h_{125}$ properties.

An interesting possibility is that $h_{125}$
might have a mixed CP nature.
Indeed,
it is known that the Standard Model (SM) CP violation and phase transition are not
enough to explain the current baryon asymmetry in the Universe. And models with extra scalars
allow for the possibility of new sources of CP violation.
Thus, there is an active experimental program to search for CP violation involving
$h_{125}$ interactions,
including:
in vector boson fusion production and its decay into four
leptons;
in vector boson fusion production using the $h_{125} \rightarrow \gamma \gamma$ channel;
a measurement of the CP properties of $h_{125} \rightarrow \tau \tau$;
and probing the CP nature of the top-Higgs Yukawa coupling in $t t h_{125}$ and $t h_{125}$ events with
$h \rightarrow b b$
decays - for reviews see for example \cite{deBlas:2019rxi,Gritsan:2022php,Angelidakis:2872699}.

A simple example
of a $h_{125}$
with a mixed CP nature
is provided by a two Higgs doublet model with
a $Z_2$ symmetry, softly-broken by a complex parameter, which provides
the single source of explicit CP violation in the scalar sector.
This is known as the complex two Higgs doublet model (C2HDM),
first introduced in \cite{Ginzburg:2002wt},
and extensively studied in 
\cite{Khater:2003wq,
ElKaffas:2007rq, Grzadkowski:2009iz, Arhrib:2010ju, Barroso:2012wz, Abe:2013qla,
Inoue:2014nva, Cheung:2014oaa, Fontes:2014xva,Grober:2017gut,Fontes:2015xva, Fontes:2015mea,
Chen:2015gaa, Chen:2017com, Muhlleitner:2017dkd, Fontes:2017zfn, Basler:2017uxn,
Aoki:2018zgq, Basler:2019iuu, Wang:2019pet, Boto:2020wyf, Cheung:2020ugr,
Altmannshofer:2020shb, Fontes:2021iue, Basler:2021kgq, Frank:2021pkc, Abouabid:2021yvw,
Fontes:2022izp,Azevedo:2023zkg,Goncalves:2023svb,Biekotter:2024ykp}. 
In particular,
as pointed out in \cite{Fontes:2014xva,Fontes:2015mea},
the C2DHM could accommodate a mostly CP-even $h_{125}tt$ coupling
in tandem with a mostly CP-odd $h_{125}bb$ coupling.
The possibility was still viable with the LHC, electron electric dipole moment
(eEDM) and flavour data available at the end of 2017 \cite{Fontes:2017zfn}.

However, recent experimental developments changed this picture.
Indeed, the analysis of the full Run2 data (both on $h_{125}$
\cite{ATLAS:2022vkf,CMS:2022dwd} and on
searches for extra scalars),
new experimental results on the eEDM \cite{ACME:2018yjb,Roussy:2022cmp}, 
a reanalysis of flavour constraints, with special emphasis on
$b \rightarrow s \gamma$,
and direct searches for CP-violation in angular correlations
of the $h_{125} \rightarrow \tau \tau$ decays \cite{CMS:2021sdq,ATLAS:2022akr},
now preclude a maximally CP-odd $h_{125}bb$ coupling in the C2HDM \cite{Biekotter:2024ykp}.
One could be led to the wrong conclusion that such a strange possibility
was loosing some of its appeal.
Part of the aim of this article is to show that that is not the case at all.
Indeed,
by adding one single extra doublet,
one can resuscitate a mostly CP-odd  $h_{125}bb$ coupling,
which, thus, should be actively sought out experimentally.

Three Higgs doublet models (3HDM) have a characteristic that enhances
the chances of success.
They relate to the possible forms of implementing natural flavour conservation (NFC)
\cite{Glashow:1976nt,Paschos:1976ay},
a mechanism of precluding flavour changing neutral couplings with scalars
(which are stringently constrained by flavour experiments such as,
for example, those in the $B - \bar{B}$ system).
Indeed, in the C2DHM there are four ways to implement NFC.
In contrast, in 3HDM there is the new possibility that each right-handed
fermion species (up-type quark, down-type quark, and charged lepton)
couples to a distinct scalar.
This is the so-called type-Z model.
The five possibilities of the 3HDM Yukawa NFC couplings are represented in Table~\ref{t:NFC},
where the scalars are denoted by $\Phi_1$, $\Phi_2$, and $\Phi_3$.
\begin{table}[h]
	\centering
	\begin{tabular}{ |c|c ccc|c| } 
		\hline
		fermion type & type-I & type-II & type-X & type-Y & type-Z \\ 
		\hline
		\hline
		up quarks ($u$) & $\Phi_3$ & $\Phi_3$ & $\Phi_3$ & $\Phi_3$ & $\Phi_3$  \\ 
		down quarks	($d$) & $\Phi_3$ & $\Phi_2$ & $\Phi_3$ & $\Phi_2$ & $\Phi_2$  \\ 
		charged leptons	($\ell$) & $\Phi_3$ & $\Phi_2$ & $\Phi_2$ & $\Phi_3$ & $\Phi_1$ \\
		\hline
	\end{tabular}
	\caption{\small Distinct possibilities for NFC in a 3HDM framework.
The first four types can also be obtained within 2HDMs but the type-Z
requires at least a 3HDM. In our convention, the scalar doublet coupling
to the up-type quarks is always labeled as $\Phi_3$.
	}
	\label{t:NFC}
\end{table}
The simplest symmetry groups that can enforce the type-Z couplings in
the 3HDM are $Z_3$ and $Z_2\times Z_2$
\cite{Alves:2020brq,Das:2022gbm,Boto:2023nyi,Cree:2011uy}, with these two models
containing all other possibilities (larger symmetry groups that have
$Z_3$ or $Z_2\times Z_2$ as subgroups).  Here, we focus on the model
with a $Z_2\times Z_2$ symmetry, with the following action on scalars,
right-handed down-type quarks ($d_R$), and right-handed
charged-leptons ($\ell_R$):
\begin{equation}\label{e:Z2Z2sym}
\begin{split}
		Z_2:\,\Phi_1 \to -\Phi_1 \, , \quad  \ell_R \to -\ell_R \,, \\
		Z'_2:\,\Phi_2 \to -\Phi_2 \, , \quad  d_R \to -d_R \,,
\end{split}
\end{equation}
(with all other SM fields invariant under the transformation)
leading to the type-Z couplings as in Table~\ref{t:NFC}.

In this  work, we study the type-Z 3HDM with a soflty-broken $Z_2\times Z_2$ symmetry,
allowing for complex parameters in the potential;
we dub this the complex 3HDM (C3HDM).
We present a parameterization of the rotations needed to go from the initial fields
to the mass eigenstates, a task complicated by the fact that the neutral scalars
have no longer a definite CP parity.
In addition, there are constraints among parameters that would seem a priori independent,
which we also explain.
Chief amongst these is an unexpected connection among the CP-odd components
of the charged leptons and the down-type quarks.
For this model we describe in detail all the theoretical and
experimental constraints that must be satisfied.

In section~\ref{sec:pot} we describe the scalar potential of the C3HDM
with $Z_2\times Z_2$ symmetry and its parameterization, while in
section~\ref{sec:Yuk} we describe the Yukawa Lagrangian. In
section~\ref{ses:constraints} we describe the constraints, both
theoretical and experimental that the model has to obey. The details
of the scan are described in section~\ref{sec:scan} and the results
are discussed in section~\ref{sec:results}. Finally, in
section~\ref{sec:conclusions} we draw our conclusions. In the
appendices some further details of the parametrization of the scalar
potential and explicit expressions for the mass matrices
are given.

\section{The scalar potential}
\label{sec:pot}

The scalar potential obeying the $Z_2\times Z_2$ symmetry in Eq.~\eqref{e:Z2Z2sym},
including soft breaking terms, is given by \cite{Botella:1994cs}
\begin{equation} \label{VNHDM}
	V = V_2 + V_4 = \mu_{ij} (\Phi_i^\dag \Phi_j) + z_{ijkl} (\Phi_i^\dag \Phi_j)(\Phi_k^\dag \Phi_l) \;,
\end{equation}
with
\begin{equation}
	V_2 =\, \mu_{11} (\Phi_1^\dag \Phi_1) + \mu_{22} (\Phi_2^\dag \Phi_2) + \mu_{33} (\Phi_3^\dag \Phi_3) + \left( \mu_{12}  (\Phi_1^\dag \Phi_2) + \mu_{13}  (\Phi_1^\dag \Phi_3) + \mu_{23}  (\Phi_2^\dag \Phi_3) + h.c.\right)
	\;, \\
\end{equation}
where $\mu_{11}$, $\mu_{22}$, $\mu_{33}$ are real, and
the complex $\mu_{12}$, $\mu_{13}$, $\mu_{23}$ parameters break the
$Z_2\times Z_2$ symmetry softly.
Also \cite{Boto:2022uwv},
\begin{equation}
\begin{split}
		V_4 =\,& V_{RI} + V_{Z_2\times Z_2} \;,\\
		V_{RI} =\,& \lambda_1 (\Phi_1^\dag \Phi_1)^2 + \lambda_2 (\Phi_2^\dag \Phi_2)^2 + \lambda_3 (\Phi_3^\dag \Phi_3)^2  + \lambda_4 (\Phi_1^\dag \Phi_1)(\Phi_2^\dag \Phi_2) + \lambda_5 (\Phi_1^\dag \Phi_1)(\Phi_3^\dag \Phi_3) \\&+ \lambda_6 (\Phi_2^\dag \Phi_2)(\Phi_3^\dag \Phi_3)  + \lambda_7 (\Phi_1^\dag \Phi_2)(\Phi_2^\dag \Phi_1) + \lambda_8 (\Phi_1^\dag \Phi_3)(\Phi_3^\dag \Phi_1) + \lambda_9 (\Phi_2^\dag \Phi_3)(\Phi_3^\dag \Phi_2) \;, \\
		V_{Z_2\times Z_2} =\,& \lambda_{10} (\Phi_1^\dag \Phi_2)^2 + \lambda_{11} (\Phi_1^\dag \Phi_3)^2 + \lambda_{12} (\Phi_2^\dag \Phi_3)^2 + h.c.
		\;,
\end{split} 
\end{equation} 
where the quartic parameters $\lambda_{10}$, $\lambda_{11}$, $\lambda_{12}$
can in general be complex, while all other parameters are real.
The potential piece $V_{RI}$ is invariant under the independent rephasings of each scalar;
the piece $V_{Z_2\times Z_2}$ is invariant under $Z_2\times Z_2$ but not under general
independent rephasings of each scalar.

To generate points of the model we need to be able to choose
the scalar masses and the vacuum expectation value (vev).
Such expressions only exist in the literature for our model in
the case where all potential parameters and the vev are real,
so we develop a parameterization for the general complex case here.

We write our doublets around a charge preserving vev as
\begin{equation}
	\Phi_i = 
	\begin{pmatrix}
		w_i^+ \\ (v_i + x_i + i\ z_i) /\sqrt{2}
	\end{pmatrix} 
	=
	\begin{pmatrix}
		w_i^+ \\ (v_i + \varphi_i) /\sqrt{2}
	\end{pmatrix} 
	\;.
\end{equation}
The vev components $v_i=|v_i| e^{-i \alpha_i}$ can be made real if
we perform a basis change that rephases each doublet by the
opposite phase of its vev component,
i.e. $\,\Phi_i' = U_{ij} \Phi_j\,$,
with $\, U = \text{diag}(e^{i\alpha_1},e^{i\alpha_2},e^{i\alpha_3})\,$.
A general basis change comes with an associated transformation in the parameters:
\begin{equation}\label{e:doub_repar}
	\mu^\prime_{ab} = U_{ai} \mu_{ij} U^\dag_{jb} 
	\qquad\text{and}\qquad 
	z^\prime_{abcd} = U_{ai} U_{ck} z_{ijkl} U^\dag_{ld} U^\dag_{jb} 
	\;,
\end{equation}
that in this case simply rephases the complex parameters:
\begin{equation}\begin{split}
		\lambda_{10} \rightarrow \lambda_{10} \cdot e^{ i\cdot 2\alpha_{12}} \;,\qquad
		\lambda_{11} \rightarrow \lambda_{11} \cdot e^{ i\cdot 2\alpha_{13}} \;,\qquad
		\lambda_{12} \rightarrow \lambda_{12} \cdot e^{ i\cdot 2\alpha_{23}} \;,\\
		\mu_{12} \rightarrow \mu_{12} \cdot e^{i\cdot \alpha_{12}} \;,\qquad
		\mu_{13} \rightarrow \mu_{13} \cdot e^{i\cdot \alpha_{13}} \;,\qquad
		\mu_{23} \rightarrow \mu_{23} \cdot e^{i\cdot \alpha_{23}} \;,
\end{split}\end{equation}
where $\alpha_{ij} = \alpha_i - \alpha_j$. 
We can therefore choose a real vev without loss of generality.

\subsection{Parameterization in terms of physical quantities}
\label{subsec:paramphys}

To relate the potential parameters with the vev and masses we need to consider
the linear and quadratic terms of the potential,
after spontaneous symmetry breaking (SSB).
We can write these in general as\footnote{Here and henceforth, bold symbols
(such as $ \boldsymbol{v}$)
refer to
vectors made out of the corresponding components (such as $v_i$).}
\begin{equation}\label{eq:V1}
	V^{(1)} = \Re(\boldsymbol{\varphi}^\dagger A \boldsymbol{v})
= \Re(\boldsymbol{x}^T A \boldsymbol{v} - i \boldsymbol{z}^T A \boldsymbol{v)}
= \boldsymbol{x}^T \Re(A\boldsymbol{v}) + \boldsymbol{z}^T \Im(A\boldsymbol{v})  \;,
\end{equation}
introducing the hermitian matrix
\begin{equation}\label{e:A}
	A_{ij} = \mu_{ij} + z_{ijkl} v_k^* v_l  \;,
\end{equation}
and as
\begin{equation}\label{eq:V2}
	\begin{split}
		V^{(2)} = 
		V^{(2)}_\text{ch} + V^{(2)}_\text{n} &= 
		( \boldsymbol{w}^+)^\dag A  \boldsymbol{w}^+
+ \frac{1}{2} \boldsymbol{\varphi}^\dag A \boldsymbol{\varphi}
+ \frac{1}{2} z_{ijkl} \varphi_i^* v_j v_k^* \varphi_l + \frac{1}{2} \Re(z_{ijkl} v_i^* \varphi_j v_k^* \varphi_l ) = \\
		&= (\boldsymbol{w}^+)^\dagger M^2_{ch} \boldsymbol{w}^+ +  
		\frac{1}{2} \begin{pmatrix}
			\boldsymbol{x}^T & \boldsymbol{z}^T
		\end{pmatrix}
		\begin{pmatrix}
			M^2_x & M^2_{xz} \\
			(M^2_{xz})^T & M^2_z 
		\end{pmatrix}
		\begin{pmatrix}
			\boldsymbol{x} \\ \boldsymbol{z}
		\end{pmatrix} = 
		\\
		&= 
		(\boldsymbol{w}^+)^\dagger M^2_{ch} \boldsymbol{w}^+ +  
		\frac{1}{2} \begin{pmatrix}
			\boldsymbol{x}^T & \boldsymbol{z}^T
		\end{pmatrix}
		M^2_n
		\begin{pmatrix}
			\boldsymbol{x} \\ \boldsymbol{z}
		\end{pmatrix}
		\;,
	\end{split}
\end{equation}
where the mass matrices are given by
\begin{equation}\label{eq:mass_matrices}\begin{split}
		M^2_{ch} &= A \;, \\
		M^2_x &= \Re(A+B + C) \;, \\
            M^2_z &= \Re(A+B - C) \;, \\
		M^2_{xz} &= -\Im(A+B+C) \;,
\end{split}\end{equation}
introducing two new matrices (hermitian and symmetric respectively)
\begin{equation}\label{e:matBC}
	B_{ij} =  z_{iklj} v_l^* v_k 
	\quad \text{and} \quad
	C_{ij} =  z_{kilj} v_l^* v_k^* 
	\;.
\end{equation}
As a result, $M^2_{ch}$ is hermitian, $M^2_x$ and $ M^2_z$ are real and symmetric,
and $M^2_{xz}$ is real, so that $M^2_n$ is a real symmetric matrix.
We include in Appendix~\ref{app:parametrization} the specific forms of the
$z_{ijkl}$ tensor and the $A$, $B$, and $C$ matrices
in terms of the potential parameters.

We impose stationarity conditions, i.e. we ensure the vev is a
stationary point of the potential.
This condition is equivalent to the tadpole condition $V^{(1)}=0$, or
\begin{equation}\label{e:stat_conditions}
	A \boldsymbol{v}=0 \,.
\end{equation}

For our model, these contain 5 independent equations, which we choose to solve for the following parameters
\small
\begin{equation}\begin{split}\label{STAT}
		\mu_{11} v_1 = -\Re(\mu_{12}) v_2 -\Re(\mu_{13}) v_3 - &v_1 \left( \lambda_1 v_1^2 + \left(\Re(\lambda_{10}) + \frac{1}{2}\lambda_4 + \frac{1}{2}\lambda_7\right) v_2^2 + \left(\Re(\lambda_{11}) + \frac{1}{2} \lambda_5 + \frac{1}{2}\lambda_8\right) v_3^2 \right ) 
		\;, \\*[1mm]
		\mu_{22} v_2 =  -\Re(\mu_{12}) v_1 -\Re(\mu_{23}) v_3 - &v_2 \left( \lambda_2 v_2^2 + \left(\Re(\lambda_{10}) + \frac{1}{2}\lambda_4 + \frac{1}{2}\lambda_7\right) v_1^2 + \left(\Re(\lambda_{12}) + \frac{1}{2}\lambda_6 + \frac{1}{2}\lambda_9\right) v_3^2 \right) 
		\;, \\*[1mm]
		\mu_{33} v_3 =  -\Re(\mu_{13}) v_1 -\Re(\mu_{23}) v_2 - &v_3 \left( \lambda_3 v_3^2 + \left(\Re(\lambda_{11}) + \frac{1}{2} \lambda_5 + \frac{1}{2}\lambda_8\right) v_1^2 + \left(\Re(\lambda_{12}) + \frac{1}{2}\lambda_6 + \frac{1}{2}\lambda_9\right) v_2^2\right) 
		\;, \\*[1mm]
		\Im(\mu_{13}) v_3 &=  -v_1\left( \Im(\lambda_{10}) v_2^2 + \Im(\lambda_{11}) v_3^2 \right) - \Im(\mu_{12}) v_2 
		\;, \\*[1mm]
		\Im(\mu_{23}) v_3 &= v_2\left( \Im(\lambda_{10}) v_1^2 - \Im(\lambda_{12})v_3^2 \right) + \Im(\mu_{12}) v_1
		\;.
	\end{split}
\end{equation}
\normalsize
These are imposed on the potential from here on.
When substituted in the mass matrices, they reveal (for a charge preserving vev) the existence of one massless charged scalar and one massless neutral scalar along the field directions $G^+ = \frac{v_i}{v} w_i^+$ and $G^0 = \frac{v_i}{v} z_i$, corresponding to the would-be Goldstone bosons from the breaking of the gauge symmetry.
We can now count the number of independent parameters. We have 24 parameters in the scalar potential,
plus the three real vevs. These are constrained by the five stationarity conditions.
Thus, we have 22 parameters. Of these, two will be reserved for $v$ and the mass of the 125GeV
scalar, leaving 20 free parameters.

Turning our attention to the mass matrices, we first need to deal with these massless directions.
These are aligned with the vev, and will be isolated if we perform a basis change that takes
it fully to the first doublet \cite{Georgi:1978ri,Donoghue:1978cj,Botella:1994cs},
i.e. if we rotate to a Higgs Basis \cite{Botella:1994cs}.

Since we are in a basis of real vevs, these may be parameterized as
\begin{equation}
\left(
\begin{array}{c}
v_1\\
v_2\\
v_3
\end{array}
\right)
=\, v\, 
\left(
\begin{array}{c}
c_{\beta_2} c_{\beta_1}\\
c_{\beta_2} s_{\beta_1}\\
s_{\beta_2}
\end{array}
\right)\, .
\label{vev:param}
\end{equation}
The Higgs basis is reached through the rotation
\begin{equation}
	R_H = R_{13}(\beta_2) R_{12}(\beta_1) = 
	\begin{pmatrix}
		c_{\beta_2} & 0 & s_{\beta_2} \\
		0 & 1 & 0 \\
		-s_{\beta_2} & 0 & c_{\beta_2} 
	\end{pmatrix}
	\begin{pmatrix}
		c_{\beta_1} & s_{\beta_1} & 0 \\
		-s_{\beta_1} & c_{\beta_1} & 0 \\
		0 & 0 & 1 
	\end{pmatrix}
	=
	\begin{pmatrix}
		c_{\beta_2}c_{\beta_1} & c_{\beta_2}s_{\beta_1} & s_{\beta_2} \\
		-s_{\beta_1} & c_{\beta_1} & 0 \\
		-s_{\beta_2}c_{\beta_1} & -s_{\beta_2}s_{\beta_1} & c_{\beta_2} 
	\end{pmatrix}\, .
\end{equation}
We could then perform a basis change on the doublets,
using $R_H$.
Such a transformation would preserve the doublet structure.
But, given that $G^0$ only has components in the $z_i$ fields,
we choose not to rotate the $x_i$ fields here,
to avoid complicating our expressions unnecessarily.
The transformations are then:
\begin{equation}\label{e:field_transf}
	\begin{pmatrix}
		x \\ z^\prime
	\end{pmatrix}
	= \begin{pmatrix}
		\one & 0 \\
		0 & R_H
	\end{pmatrix}
	\begin{pmatrix}
		x \\ z
	\end{pmatrix}
	\;,
	\qquad \qquad
	w^{+\prime} = R_H w^+
	\; ,
\end{equation}
\begin{equation}
 R_H M_{ch}^2 R_H^T =\begin{pmatrix}
     0 & 0 & 0  \\
    0 &\multicolumn{2}{c}{\multirow{2}{*}{ $ M_{ch}^{2\prime}$}}\\ 
    0  &
    \end{pmatrix}\;, 
\qquad  \quad  
\begin{pmatrix}
		\one & 0 \\
		0 & R_H
	\end{pmatrix}M_{n}^2\begin{pmatrix}
		\one & 0 \\
		0 & R_H^T
	\end{pmatrix}=\begin{pmatrix}
\multicolumn{3}{c}{\multirow{3}{*}{ $ M_{x}^{2}$}}  & 0 & \multicolumn{2}{c}{\multirow{3}{*}{ $ M_{xz}^{2\prime}$}}\\ 
      &  &  & 0 &  &   \\
           &  &  & 0 &  &   \\
     0 & 0 & 0 & 0 & 0 & 0 \\
    \multicolumn{3}{c}{\multirow{2}{*}{ $ M_{xz}^{2\prime T}$}}  & 0 & \multicolumn{2}{c}{\multirow{2}{*}{ $ M_z^{2\prime}$}}\\ 
     &  &  & 0 &  &   
    \end{pmatrix}
		\;.
\end{equation}

The resulting hermitian/symmetric mass matrices
${M_{ch}^{2\prime}}$/$M_n^{2\prime}$ are written in
Appendix \ref{app:mass_matrices}.
These can now be diagonalized using unitary/orthogonal transformations. 
However, looking at these matrices,
we find that the 12 entries in the real part of $M^{2\prime}_{ch}$,
the upper-left $3\times 3$,
and the lower-right $2\times 2$ blocks of $M^{2\prime}_n$,
are regulated by the 15 real parameters 
\begin{equation}\label{e:re_parameteres}
	\lambda_1, ..., \lambda_9, \Re(\lambda_{10}), \Re(\lambda_{11}), \Re(\lambda_{12}), \Re(\mu_{12}), \Re(\mu_{13}), \Re(\mu_{23})
	\,,
\end{equation}
while the 4 imaginary parameters 
\begin{equation}\label{e:im_parameters}
	\Im(\lambda_{10}), \Im(\lambda_{11}), \Im(\lambda_{12}), \Im(\mu_{12})
\end{equation}
regulate the 7 entries in the imaginary part of $M^{2\prime}_{ch}$ and the remaining upper-right $3\times 2$ block of $M^{2\prime}_n$.
This means that the neutral mass matrix is not a general real
symmetric matrix and we cannot choose arbitrary eigenvalues
and orthogonal eigenvectors for it.
Since we do not know how to generate eigenvectors and eigenvalues
that produce matrices of this form, we consider a general orthogonal rotation,
expecting to obtain $7-4=3$ constraints at the end of the parameterization.

For our 2 charged and 5 neutral scalars, this is
\begin{equation}\label{mass_eigenstates}
	\Big(H_1^+ \; H_2^+\Big)^T = W\;\Big(w_2^{+\prime}\; w_3^{+\prime}\Big)^T  \;, \qquad \quad \Big(h_1\;h_2\;h_3\;h_4\;h_5\Big)^T = R 
	\;\Big(x_1\;x_2\;x_3\;z_2^\prime\; z_3^\prime\Big)^T \;,
\end{equation}
such that
\begin{equation} \label{e:mass_equations}
	M_{ch}^{2\prime} = W^\dag 
	\begin{pmatrix}
		m_{H_1^\pm}^2 & 0 \\ 0 & m_{H_2^\pm}^2
	\end{pmatrix}
	W
	\;,
	\qquad \quad
	M_{n}^{2\prime} = R^T 
	\begin{pmatrix}
		m_{h_1}^2 & 0 & 0 & 0 & 0 \\ 
		0 & m_{h_2}^2 & 0 & 0 & 0 \\ 
		0 & 0 & m_{h_3}^2 & 0 & 0 \\ 
		0 & 0 & 0 & m_{h_4}^2 & 0 \\ 
		0 & 0 & 0 & 0 & m_{h_5}^2 
	\end{pmatrix}
	R
	\;,
\end{equation}
where $m_{h_i}^2$/$m_{H_1^\pm}^2$ are the squared masses of the neutral/charged scalars.
In Eqs.~\eqref{mass_eigenstates}, the states are listed in order of increasing mass. In this article,
we take the lightest neutral scalar to be the state found at LHC ($h_1=h_{125}$),
as our main goal is to make the point that it can have a large pseudoscalar coupling to the
bottom quark. We can parameterize the unitary $W$ using two angles $\theta$ and $\varphi$, as
\begin{eqnarray}
	W &=& 
	\begin{pmatrix}
		c_\theta e^{i \varphi} & s_\theta e^{-i \varphi} \\*[1mm]
		-s_\theta e^{i \varphi} & c_\theta e^{-i \varphi}
	\end{pmatrix}
\nonumber\\
&&
\Rightarrow\ 
	W^\dag
	\begin{pmatrix}
		m_{H_1^\pm}^2 & 0 \\*[2mm] 0 & m_{H_2^\pm}^2
	\end{pmatrix}
	W
	= 
	\begin{pmatrix}
		m_{H_1^\pm}^2 c_\theta^2 + m_{H_2^\pm}^2 s_\theta^2 & (m_{H_1^\pm}^2-m_{H_2^\pm}^2) s_\theta c_\theta e^{-2 i \varphi} \\*[2mm]
		(m_{H_1^\pm}^2-m_{H_2^\pm}^2) s_\theta c_\theta e^{2 i \varphi} & m_{H_1^\pm}^2 s_\theta^2 + m_{H_2^\pm}^2 c_\theta^2
	\end{pmatrix}
	\;,
\end{eqnarray}
and $R$ as some product of the ten rotations $R_{ij}$ in five dimensions.

The equality of matrices in \eqref{e:mass_equations} contains 19 independent real equations, linear in the squared masses, and in the potential parameters $\mu_{ij}$ and $\lambda_i v^2$, which we solve for.
The $19=12+4+3$ equations can be solved for 12 of the 15 parameters in Eq.~\eqref{e:re_parameteres}
and the 4 parameters in Eq.~\eqref{e:im_parameters}, yielding
\begin{equation}\label{eq:L1}
	\lambda_1 v^2 = \frac{1}{2 c_{\beta_1}^2 c_{\beta_2}^2  }\left[ t_{\beta_1} \Re(\mu_{12}) + \frac{t_{\beta_2}}{c_{\beta_1}} \Re(\mu_{13}) + R_{i1}^2 m_{h_i}^2 \right]
	\;,
\end{equation}
\begin{equation}\label{eq:L2}
	\lambda_2 v^2 = \frac{1}{2 s_{\beta_1}^2 c_{\beta_2}^2  }\left[ \frac{1}{t_{\beta_1}} \Re(\mu_{12}) + \frac{t_{\beta_2}}{s_{\beta_1}} \Re(\mu_{23}) + R_{i2}^2 m_{h_i}^2 \right]
	\;,
\end{equation}
\begin{equation}\label{eq:L3}
	\lambda_3 v^2 = \frac{1}{2 s_{\beta_2}^2  }\left[ \frac{c_{\beta_1}}{t_{\beta_2}} \Re(\mu_{13}) + \frac{s_{\beta_1}}{t_{\beta_2}} \Re(\mu_{23}) + R_{i3}^2 m_{h_i}^2 \right]
	\;,
\end{equation}
\begin{equation}\label{eq:L4}
	\lambda_4 v^2 = -\lambda_7 v^2 - 2 \Re(\lambda_{10}) v^2 + \frac{1}{c_{\beta_1} s_{\beta_1} c_{\beta_2}^2  }\left[ -\Re(\mu_{12}) + R_{i1} R_{i2} m_{h_i}^2 \right]
	\;,
\end{equation}
\begin{equation}\label{eq:L5}
	\lambda_5 v^2 = -\lambda_8 v^2 - 2 \Re(\lambda_{11}) v^2 + \frac{1}{c_{\beta_1} c_{\beta_2} s_{\beta_2}  }\left[ -\Re(\mu_{13}) + R_{i1} R_{i3} m_{h_i}^2 \right]
	\;,
\end{equation}
\begin{equation}
	\lambda_6 v^2 = -\lambda_9 v^2 - 2 \Re(\lambda_{12}) v^2+ \frac{1}{ s_{\beta_1} c_{\beta_2} s_{\beta_2}  }\left[ -\Re(\mu_{23}) + R_{i2} R_{i3} m_{h_i}^2 \right]
	\;,
\end{equation}
\begin{equation}
	\begin{split}
		\lambda_7 v^2 = - 2 \Re(\lambda_{10}) v^2 + \frac{1}{c_{\beta_1} s_{\beta_1} c_{\beta_2}^2  }\Big[ -2\Re(\mu_{12}) 
		+2\Big( 
		-c_{\beta_1}s_{\beta_1}(c_\theta^2 m_{H_1^\pm}^2 + s_\theta^2 m_{H_2^\pm}^2) \\
		+ c_{\beta_1}s_{\beta_1} s_{\beta_2}^2 (s_\theta^2 m_{H_1^\pm}^2 + c_\theta^2 m_{H_2^\pm}^2)
		+c_\theta s_\theta s_{\beta_2} (1-2 s_\varphi^2)(1-2 s_{\beta_1}^2)(m_{H_2^\pm}^2-m_{H_1^\pm}^2)
		\Big) \Big]
		\;,
	\end{split}
\end{equation}
\normalsize
\begin{equation}
	\begin{split}
		\lambda_8 v^2 = - 2 \Re(\lambda_{11}) v^2& + \frac{1}{c_{\beta_1} s_{\beta_2} c_{\beta_2}  }\Big[ -2\Re(\mu_{13}) \\
		+2 c_{\beta_2}\Big( 
		-c_{\beta_1}s_{\beta_2}&(s_\theta^2 m_{H_1^\pm}^2 + c_\theta^2 m_{H_2^\pm}^2)
		+c_\theta s_\theta s_{\beta_1} (1-2 s_\varphi^2)(m_{H_2^\pm}^2-m_{H_1^\pm}^2)
		\Big) \Big]
		\;,
	\end{split}
\end{equation}
\normalsize
\begin{equation}
	\begin{split}
		\lambda_9 v^2 = - 2 \Re(\lambda_{12}) v^2& + \frac{1}{s_{\beta_1} s_{\beta_2} c_{\beta_2}  }\Big[ -2\Re(\mu_{23}) \\
		+2 c_{\beta_2}\Big( 
		-s_{\beta_1}s_{\beta_2}&(s_\theta^2 m_{H_1^\pm}^2 + c_\theta^2 m_{H_2^\pm}^2)
		-c_\theta s_\theta c_{\beta_1} (1-2 s_\varphi^2)(m_{H_2^\pm}^2-m_{H_1^\pm}^2)
		\Big) \Big]
		\;,
	\end{split}
\end{equation}
\normalsize
\begin{equation}
	\Re(\lambda_{10}) v^2 = \frac{-1}{2 c_{\beta_2}^2} \left[
	\frac{1}{s_{\beta_1}c_{\beta_1}} \Re(\mu_{12}) + m_{h_i}^2\left( R_{i4}^2 + s_{\beta_2}\frac{2 c_{\beta_1}^2-1}{s_{\beta_1}c_{\beta_1}} R_{i4}R_{i5} - R_{i5}^2 s_{\beta_2}^2 \right)
	\right]
	\;,
\end{equation}
\begin{equation}
	\Re(\lambda_{11}) v^2 = \frac{-1}{2 c_{\beta_1} s_{\beta_2}} \left[
	\frac{1}{c_{\beta_2}} \Re(\mu_{13}) + m_{h_i}^2 R_{i5}\left( R_{i4} s_{\beta_1} + R_{i5} c_{\beta_1} s_{\beta_2} \right)
	\right]
	\;,
\end{equation}
\begin{equation}
	\Re(\lambda_{12}) v^2 = \frac{-1}{2 s_{\beta_1} s_{\beta_2}} \left[
	\frac{1}{c_{\beta_2}} \Re(\mu_{23}) + m_{h_i}^2 R_{i5}\left( R_{i5} s_{\beta_1}s_{\beta_2} - R_{i4} c_{\beta_1} \right)
	\right]
	\;,
\end{equation}
\begin{equation}\label{eq:L10I}
	\Im(\lambda_{10}) v^2 = \frac{1}{2 c_{\beta_1}s_{\beta_1}c_{\beta_2}^2} \left[
	- \Im(\mu_{12}) + m_{h_i}^2 R_{i1} \left( R_{i5} s_{\beta_1}s_{\beta_2} - R_{i4}c_{\beta_1}  \right)
	\right]
	\;,
\end{equation}
\begin{equation}\label{eq:L11I}
	\Im(\lambda_{11}) v^2 = \frac{s_{\beta_1}}{2 c_{\beta_1}s_{\beta_2}^2} \left[
	\Im(\mu_{12}) - m_{h_i}^2 R_{i1} \left( R_{i4} c_{\beta_1} + R_{i5} \frac{s_{\beta_2}(1+c_{\beta_1}^2)}{s_{\beta_1}}  \right)
	\right]
	\;,
\end{equation}
\begin{equation}\label{eq:L12I}
	\Im(\lambda_{12}) v^2 = \frac{c_{\beta_1}}{2 s_{\beta_1}s_{\beta_2}^2} \left[
	- \Im(\mu_{12}) + m_{h_i}^2 \left( R_{i1}(R_{i4}c_{\beta_1} -R_{i5}s_{\beta_1}s_{\beta_2} ) - \frac{2 s_{\beta_2}}{c_{\beta_1}} R_{i2}R_{i5} \right)
	\right]
	\;,
\end{equation}
\begin{equation}\label{eq:mu12I}
	\Im(\mu_{12}) = 4 c_\varphi s_\varphi c_\theta s_\theta s_{\beta_2} (m_{H_2^\pm}^2-m_{H_1^\pm}^2) + m_{h_i}^2 R_{i1}(R_{i4}c_{\beta_1}-R_{i5}s_{\beta_1}s_{\beta_2})
	\;,
\end{equation}
with 3 additional constraints on the neutral mass matrix:
\begin{equation}\label{e:X1i}
	m_{h_i}^2 \left[ R_{i5} c_{\beta_2}(R_{i1}s_{\beta_1}-R_{i2}c_{\beta_1}) - R_{i3}R_{i4} \right] = X_{1i} m_{h_i}^2 = 0
	\;,
\end{equation}
\begin{equation}\label{e:X2i}
	m_{h_i}^2 R_{i5} \frac{ c_{\beta_2}(R_{i1}c_{\beta_1}+R_{i2}s_{\beta_1}) - R_{i3} s_{\beta_2} }{s_{\beta_2}}  = X_{2i} m_{h_i}^2  = 0
	\;,
\end{equation}
\begin{equation}\label{e:X3i}
	m_{h_i}^2 \frac{ R_{i4} (R_{i1}c_{\beta_1}-R_{i2}s_{\beta_1}) - R_{i5} s_{\beta_2}(R_{i1}s_{\beta_1}+R_{i2}c_{\beta_1}) }{s_{\beta_1}}  = X_{3i} m_{h_i}^2  = 0
	\;.
\end{equation}

We have defined, in
Eqs.~\eqref{e:X1i}-\eqref{e:X3i},
a $3\times 5$ matrix $X$, which is used to determine three of the five masses,
with the other two being independent. 
Assuming that, 
given $m_{h_1}^2$ and $m_{h_2}^2$,
the system can be solved for $m_{h_3}^2$, $m_{h_4}^2$ and $m_{h_5}^2$.
We have
\begin{equation}\label{e:m3sq}
	\begin{split}
		m_{h_3}^2 = 
		-\frac{ \sum_{i=1}^2 (X_{1i}X_{24}X_{35}-X_{1i}X_{25}X_{34}-X_{14}X_{2i}X_{35}+X_{14}X_{25}X_{3i}+X_{15}X_{2i}X_{34}-X_{15}X_{24}X_{3i}) m_{h_i}^2}{X_{13}X_{24}X_{35}-X_{13}X_{25}X_{34}-X_{14}X_{23}X_{35}+X_{14}X_{25}X_{33}+X_{15}X_{23}X_{34}-X_{15}X_{24}X_{33}} 
		\;,
	\end{split}
\end{equation}
\begin{equation}\label{e:m4sq}
	m_{h_4}^2 = - \frac{1}{X_{24}X_{35}-X_{25}X_{34}} \sum_{i=1}^3 (X_{2i}X_{35}- X_{3i} X_{25}) m_{h_i}^2
	\;,
\end{equation}
\begin{equation}\label{e:m5sq}
	m_{h_5}^2 = -\frac{1}{X_{35}}\sum_{i=1}^4  X_{3i}  m_{h_i}^2
	\;.
\end{equation}

The non-generality of the mass matrices is similar to what happens
in the C2HDM \cite{Fontes:2017zfn}, where there is a single constraint.
However, the presence of one single constraint on the C2HDM makes it
easily solvable for one of the rotation angles, unlike in our C3HDM,
where the situation is much more complicated.
This increases the difficulty of generating parameter space points,
as we cannot choose all of the masses, and we can get negative
square masses from expressions
\eqref{e:m3sq}-\eqref{e:m5sq} that need to be discarded.

\subsection{The Real Limit}
\label{subsec:real}

The real limit is attained by making all of the imaginary parts of potential
parameters zero, i.e. $\Im(\mu_{12})$, $\Im(\mu_{13})$, $\Im(\mu_{23})$,
$\Im(\lambda_{10})$, $\Im(\lambda_{11})$, $\Im(\lambda_{12})$ = 0. 
The first 3 stationarity conditions in Eq.~\eqref{STAT} are unaltered,
while the other 2 become trivial .
The mass matrices are fully real in this case,
and there is no mixing
between scalars $x_i$ and pseudoscalars $z_i$,
so we must simplify our rotation matrices.
We set $\varphi=0$ for the charged scalars,
while the neutral scalars' rotation matrix takes the simpler
form\footnote{Here we are using a specific parametrization for $R$,
which will be specified in detail in Eq.~\eqref{eq:Rspecific}
of section~\ref{sec:scan} below.}
\begin{equation}
	\begin{split}
		R = R_z R_x =  R_{45}(\alpha_{45}) \,\cdot 
R_{23}(\alpha_{23}) R_{13}(\alpha_{13}) R_{12}(\alpha_{12}) \,.
	\end{split}
\end{equation}

Turning to the parametrization coming from the mass matrices,
the three constraints \eqref{e:X1i}-\eqref{e:X3i},
and the equations for the imaginary parts of parameters
Eqs.~\eqref{eq:L10I}-\eqref{eq:mu12I} all become trivial.
The rest of the parametrization should then agree with
the parametrization of the real model, with the correspondence
between our angles and the angles in \cite{Das:2019yad,Boto:2022uwv},
given by
\begin{equation}
	\begin{split}
		&\alpha_{12} = \alpha_1 \;,\qquad
		\alpha_{13} = \alpha_2 \;,\qquad
		\alpha_{23} = \alpha_3 \;,\\
		&\qquad\qquad \alpha_{45} = -\gamma_1 \;, \qquad
		\theta = -\gamma_2 \;.
	\end{split}
\end{equation}

\section{\label{sec:Yuk}The Yukawa Lagrangian}

\subsection{Higgs-Fermion Couplings in Generic NFC}

We are interested in looking at the couplings between the fermions and the neutral scalars.
These come from the Yukawa Lagrangian, which,
assuming that right-handed fermions of each electric charge couple only to one Higgs doublet
- indicated henceforth by $\Phi_d$, $\Phi_u$ and $\Phi_\ell$ -
can be written as
\begin{equation}\label{LY}
- \mathcal{L}_\text{Yukawa} = \overline{Q}_L \Gamma \Phi_d n_R
+\overline{Q}_L \Delta \tilde{\Phi}_u p_R
+\overline{L}_L Y \Phi_\ell \ell_R + h.c. 
\;,
\end{equation}
where $Q_L = (p_L \; n_L)^T$, $L_L = (\nu_L \; \ell_L)^T$,
while $n_R$, $p_R$, and $\ell_R$ are,
respectively,
right-handed down-type, up-type, and charged lepton fields,
written in a weak basis.
For simplicity, we do not include right-handed neutrinos, rendering neutrinos massless.
After SSB, the fermions gain mass terms
\begin{eqnarray}
- \mathcal{L}_\text{Yukawa}
&=& 
\frac{v_d}{\sqrt{2}} \overline{n}_L \Gamma n_R + (\overline{p}_L \; \overline{n}_L) \Gamma \begin{pmatrix}
	\varphi_d^+ \\ (x_d+iz_d)/\sqrt{2}
\end{pmatrix} n_R
\nonumber\\*[2mm]
&&
+\, \frac{v_u}{\sqrt{2}} \overline{p}_L \Delta p_R + (\overline{p}_L \; \overline{n}_L) \Delta \begin{pmatrix}
	(x_u-iz_u)/\sqrt{2} \\ -\varphi_u^-
\end{pmatrix} p_R
\nonumber\\*[2mm]
&&
+\, \frac{v_\ell}{\sqrt{2}} \overline{\ell}_L Y \ell_R + (\overline{\nu}_L \; \overline{\ell}_L) Y \begin{pmatrix}
	\varphi_\ell^+ \\ (x_\ell+i z_\ell)/\sqrt{2}
\end{pmatrix} \ell_R
+ h.c. \;.
\end{eqnarray}

Just as in the SM, fermions of each flavour and chirality
can be reparameterized by unitary transformations
\begin{equation}
	\begin{split}
		p_L = U_L^p u_L\;, \qquad p_R = U_R^p u_R\;, & \qquad n_L = U_L^n d_L\;, \qquad n_R = U_R^n d_R\;, \\*[1mm]
		\ell_L = U_L^\ell \tilde{\ell}_L\;, \qquad \ell_R = & U_R^\ell \tilde{\ell}_R\;, \qquad \nu_L = U_L^\nu \tilde{\nu}_L \;,
	\end{split}
\end{equation}
which can be used to diagonalize the mass terms
\begin{equation}
	\begin{split}
		\frac{v_d}{\sqrt{2}} \Gamma = M_n\;,& \qquad (U_L^n)^\dagger M_n U_R^n = D_d = \text{diag}(m_{di})\;, \\
		\frac{v_u}{\sqrt{2}} \Delta = M_p\;,& \qquad (U_L^p)^\dagger M_p U_R^p = D_u = \text{diag}(m_{ui})\;,  \\
		\frac{v_\ell}{\sqrt{2}} Y = M_\ell\;,& \qquad (U_L^\ell)^\dagger M_\ell U_R^\ell = D_\ell = \text{diag}(m_{\ell i})\;. 
	\end{split}
\end{equation}
Since there are no right-handed neutrino fields,
there is no neutrino mass matrix,
and we can choose $U_L^\nu = U_L^\ell$ so that $(\tilde{\nu}_L \; \tilde{\ell}_L)^T$
is also an $SU(2)$ doublet.
Thus, we can simply reparameterize the leptons,
assume the matrix $Y$ was diagonal from the beginning, and drop the tildes.
This is not the case of the quarks, where the matrix $V = U_L^{p\dagger}U_L^n$
cannot be chosen to be the identity, since $U_L^p$ and $U_L^n$ are both fixed
by the bi-diagonalization of the respective mass matrices.
$V$ is the Cabibbo-Kobayashi-Maskawa \cite{Cabibbo:1963yz,Kobayashi:1973fv}
matrix (CKM matrix).

The Lagrangian has then terms of the form
\begin{equation}
	\mathcal{L}_\text{Yukawa} = \mathcal{L}_{ff} + \mathcal{L}_{\Phi ff} 
	\;,
\end{equation}
where
\begin{equation}
-\mathcal{L}_{ff} = \overline{d}_L D_d d_R 
+ \overline{u}_L D_u u_R 
+ \overline{\ell}_L D_\ell \ell_R + h.c. 
= 
\overline{d} D_d d + \overline{u} D_u u + \overline{\ell} D_\ell \ell\, , 
\end{equation}
are the fermion mass terms, and $\mathcal{L}_{\Phi ff}$ contains the scalar-fermion interactions
\small
\begin{equation}
	\begin{split}
		- \mathcal{L}_{\Phi ff} = 
		(\overline{p}_L \; \overline{n}_L) \Gamma \begin{pmatrix}
			\varphi_d^+ \\ (x_d+iz_d)/\sqrt{2}
		\end{pmatrix} n_R
		+ (\overline{p}_L \; \overline{n}_L) &\Delta \begin{pmatrix}
			(x_u-iz_u)/\sqrt{2} \\ -\varphi_u^-
		\end{pmatrix} p_R 
		\\
		+\, (\overline{\nu}_L \; \overline{\ell}_L) Y \begin{pmatrix}
			\varphi_\ell^+ \\ (x_\ell+i z_\ell)/\sqrt{2}
		\end{pmatrix}& \ell_R
		+ h.c.
		\\
		=
		 \left(\overline{u}_L U_L^{p \dagger} \;\;\; \overline{d}_L U_L^{n \dagger}\right) \Gamma \begin{pmatrix}
			\varphi_d^+ \\ (x_d+iz_d)/\sqrt{2}
		\end{pmatrix} U_R^n d_R
		+ \left(\overline{u}_L U_L^{p \dagger} \;\;\; \overline{d}_L U_L^{n \dagger} \right) &\Delta \begin{pmatrix}
			(x_u-iz_u)/\sqrt{2} \\ -\varphi_u^-
		\end{pmatrix} U_R^p u_R 
		\\
		+\, (\overline{\nu}_L \; \overline{\ell}_L) Y \begin{pmatrix}
			\varphi_\ell^+ \\ (x_\ell+i z_\ell)/\sqrt{2}
		\end{pmatrix}& \ell_R
		+ h.c.
		\\
		=
		 \left(\overline{u}_L V \;\;\; \overline{d}_L \right) \frac{\sqrt{2} D_d}{v_d} \begin{pmatrix}
			\varphi_d^+ \\ (x_d+iz_d)/\sqrt{2}
		\end{pmatrix} d_R
		+ \left(\overline{u}_L \;\;\; \overline{d}_L V^\dagger \right) &\frac{\sqrt{2} D_u}{v_u} \begin{pmatrix}
			(x_u-iz_u)/\sqrt{2} \\ -\varphi_u^-
		\end{pmatrix} u_R 
		\\*[1mm]
		+\, (\overline{\nu}_L \; \overline{\ell}_L) \frac{\sqrt{2} D_\ell}{v_\ell} \begin{pmatrix}
			\varphi_\ell^+ \\ (x_\ell+i z_\ell)/\sqrt{2}
		\end{pmatrix}& \ell_R
		+ h.c.
	\end{split}
\end{equation}
\normalsize
where the CKM matrix $V$
is involved in the interactions with charged scalars.
The interactions with neutral scalars are
\begin{equation} \label{Lh0ff}
	-\mathcal{L}_{\xi ff} = 
	\frac{m_{\ell i}}{v_\ell} \overline{\ell}_i (x_\ell + i \gamma_5 z_\ell) \ell_i 
	+\frac{m_{di}}{v_d} \overline{d}_i (x_d + i \gamma_5 z_d) d_i 
	+\frac{m_{ui}}{v_u} \overline{u}_i (x_u - i \gamma_5 z_u) u_i 
	\; .
\end{equation}

We now need to write this Lagrangian using the scalars' mass eigenstates.
In general, the neutral scalar mass matrix is diagonalized by some orthogonal
transformation $Q$:
\begin{equation}
	\begin{pmatrix}
		\xi_1 \\ ...  \\ \xi_{2N}
	\end{pmatrix}
	=
	Q
	\begin{pmatrix}
		x_1 \\ ... \\ x_N \\ z_1 \\ ... \\ z_N
	\end{pmatrix}
	\quad \iff \quad
	x_i = Q^T_{ij}\; \xi_j = Q_{ji} \;\xi_j \quad,\quad z_i = Q^T_{N+i,j}\; \xi_j = Q_{j,N+i} \;\xi_j
	\;,
\end{equation}
where $\xi_1 = G^0$ is the pseudo-Goldstone boson absorbed by the Z boson,
and $i \in \{ 1, \dots, N \}$, while $j \in \{ 1, \dots, 2N \}$.
With this substitution, we can write \eqref{Lh0ff} as
\begin{equation}\begin{split}
		-\mathcal{L}_{\xi ff} = 
		\sum_f \sum_{j=1}^{2N}  \frac{m_f}{v} \overline{f} \frac{v}{v_f}\left( Q_{jf} \pm i \gamma_5 Q_{j,N+f}\right) f\; \xi_j 
		=
		\sum_f \sum_{j=1}^{2N}  \frac{m_f}{v} \overline{f} \left( c^e_{\xi_j ff} + i\gamma_5 c^o_{\xi_j ff} \right) f\; \xi_j 
		\;,
\end{split}\end{equation}
where we defined 
\begin{equation}\label{h0ff_couplings}
	\;\; c^e_{\xi_j ff} + i\gamma_5 c^o_{\xi_j ff} = \frac{v}{v_f}\left( Q_{jf} \pm i \gamma_5 Q_{j,N+f}\right)
	\;.
\end{equation}
The plus sign is used for leptons and down-type quarks, and the minus sign for up-type quarks.
To obtain these Feynman
rules we have implemented the model in the
\texttt{FeynMaster} \cite{Fontes:2019wqh,Fontes:2021iue} software
package.\footnote{The complete and consistent set of all
Feynman Rules for this model
may be found at the url
\url{https://porthos.tecnico.ulisboa.pt/~romao/Work/arXiv/C3HDMZ2xZ2/}.}

\subsection{Higgs-Fermion Couplings in the \texorpdfstring{$Z_2\times Z_2$}{} - 3HDM}

The Higgs-fermion couplings for our model are obtained from the rotation matrices
in Appendix~\ref{a:rotation_matrices};
the matrix $Q$ being defined in \eqref{matrixQ},
where, as stated, we choose the first position for the field $\xi_1 = G^0$.

From \eqref{h0ff_couplings}, and assuming that the Goldstone is $\xi_1 = G^0$,
and that the 125 GeV Higgs is $h_{125} = h_1 = \xi_2$, we have
\begin{equation}
	c^e_{h_{125}ff } + i\gamma_5 c^o_{h_{125}ff} = \frac{v}{v_f}\left( Q_{2,f} \pm i \gamma_5 Q_{2,3+f}  \right)\;,
\end{equation}
where $f$ stands for the index of the doublet that couples to the fermion in question in the Yukawa Lagrangian \eqref{LY}, and the $\pm$ is for leptons/down-type quarks, and up-type quarks, respectively.  We then have, for the different possibilities of $f$
\begin{equation}
	c^e_{h_{125}ff} =  
	\frac{R_{11}}{c_{\beta_2} c_{\beta_1}} \;,\;
	\frac{R_{12}}{c_{\beta_2} s_{\beta_1}} \;,\;
	\frac{R_{13}}{s_{\beta_2} } \;,\;
	\qquad \text{for} \quad f=1,2,3 \;,
\end{equation}
and
\begin{equation}
	c^o_{h_{125}ff} = 
	\pm\frac{-R_{14}s_{\beta_1}-R_{15}c_{\beta_1}s_{\beta_2}}{c_{\beta_2} c_{\beta_1}} \;, \;
	\pm\frac{R_{14}c_{\beta_1}-R_{15}s_{\beta_1}s_{\beta_2}}{c_{\beta_2} s_{\beta_1}} \;,\;
	\pm  \frac{R_{15} c_{\beta_2}}{s_{\beta_2}}
	\qquad \text{for} \quad f=1,2,3 \;.
\end{equation}
In a type-Z model, each possibility $f=1,2,3$ will correspond to the
coupling of the $h_{125}$ to fermions of a different charge,
respectively, charged lepton, down-type quark and up-type quark,
as seen in Table~\ref{t:NFC}.
%
%

\subsection{Inverting the Higgs Couplings}
\label{s:HiggsCouplings}

We are interested in inverting the relation between observable couplings of the Higgs with fermions and gauge bosons, and the parameters of the model that regulate them. 
The trilinear couplings of gauge boson to neutral scalars arise from the action of the covariant derivative on the doublets. 
In particular, they couple only to the scalar aligned with the vev $\,x_1^\prime = (R_H)_{1i}x_i = (R_H)_{1i} Q^T_{ij} \xi_j\,$. Remembering that we identify $\xi_2= h_1$ with the $h_{125}$, we have 
\begin{equation}\label{e:kV}
	\begin{split}
		\mathcal{L} \supseteq &  
		\left(\frac{g^2 v}{2} W_\mu^+ W_\mu^- +  \frac{g^2 v}{4 c_W^2} Z_\mu^2 \right) x_1^\prime \supseteq
		\left(\frac{g^2 v}{2} W_\mu^+ W_\mu^- + \frac{g^2 v}{4 c_W^2} Z_\mu^2 \right) 
		(R_H)_{1i} Q_{2i} \xi_2 =
		\mathcal{L}^{(SM)}_{hVV} \kappa_V \,, \\*[3mm]
		&\qquad\qquad\qquad\qquad\qquad \kappa_V = \left(\begin{pmatrix}
			R_H & 0 \\ 0 & R_H
		\end{pmatrix} Q^T\right)_{12} = (R_H)_{1i} Q_{2i}  = Q_{2i} \hat{v}_i
		\,.
	\end{split}
\end{equation}

The couplings with fermions have been obtained above. 
We rewrite them here, for the coupling with $\xi_2= h_1 = h_{125}$,
with the correspondence $i=1,2,3=\tau,b,t$,
and absorbing the negative sign in the odd top coupling for
simplicity\footnote{The definition of $c_3^o=c_t^o$ made here has a sign difference
with respect to the definition for the C2HDM in \cite{Fontes:2017zfn,Biekotter:2024ykp}.
We use here Eq.~\eqref{e:ceco} because it leads to more symmetric equations
in what follows below.}
\begin{equation}\label{e:ceco}
	\begin{split}
		c^e_i = c^e_{h_{125}ii} = \frac{Q_{2i}}{\hat{v}_i}  \,,\quad
		c^o_i = (-1)^{\delta_{i3}} c^o_{h_{125}ii}  = \frac{Q_{2,3+i}}{\hat{v}_i}  \,.
	\end{split}
\end{equation} 
We will also denote the absolute value of these couplings as $\,k^2_i = (c^e_i)^2 + (c^o_i)^2$.

The seven couplings $\kappa_V$, $c^e_i$, $c^o_i$ are a function of the vev direction $\hat{v}_i$ - regulated by 2 angles, since $\hat{v}_i\hat{v}_i=1$ - and of the second line of the rotation matrix $Q_{2i}$. 
This is the line of an orthogonal matrix, so it obeys $Q_{2 i} Q_{2 i} + Q_{2,3+i} Q_{2,3+i} = 1 $, and an additional constraint $Q_{2,3+i} \hat{v}_i = 0$, so it is regulated by 4 parameters.
Our model should then predict 1 constraint between these 7 observables, as they are regulated by 6 parameters.

We can write the full set of relations between parameters and couplings - using \eqref{e:ceco} to replace $Q_{2i}$, $Q_{2,3+i}$ - as
\begin{align}
	Q_{2i} &= c^e_i \hat{v}_i \,, \label{e:e1}  \\
	Q_{2,3+i} &= c^o_i \hat{v}_i \,, \label{e:e2} \\
	c^e_i \hat{v}_i^2 &= \kappa_V \,, \label{e:e3} \\
	c^o_i \hat{v}_i^2 &= 0 \,, \label{e:e4} \\
	\hat{v}_i\hat{v}_i&=1 \,, \label{e:e5} \\
	(c^e_i)^2 \hat{v}_i^2 + (c^o_i)^2 &\hat{v}_i^2 = k^2_i \hat{v}_i^2  = 1 \,. \label{e:e6} 
\end{align}
We use (\ref{e:e4}-\ref{e:e6}) to solve for the vev
\begin{equation}\label{e:vev_sol}
	\begin{split}
		\hat{v}_1^2 = \frac{\Delta_2 c^o_3 - \Delta_3 c^o_2}{\Delta_1 c^o_{23}+\Delta_2 c^o_{31}+\Delta_3 c^o_{12}} \,,\\*[1mm]
		\hat{v}_2^2 = \frac{\Delta_3 c^o_1 - \Delta_1 c^o_3}{\Delta_1 c^o_{23}+\Delta_2 c^o_{31}+\Delta_3 c^o_{12}} \,,\\*[1mm]
		\hat{v}_3^2 = \frac{\Delta_1 c^o_2 - \Delta_2 c^o_1}{\Delta_1 c^o_{23}+\Delta_2 c^o_{31}+\Delta_3 c^o_{12}} \,,\\
	\end{split}
\end{equation}
where $\Delta_i= \kappa^2_i-1$, and $c^o_{ij} = c^o_i-c^o_j$. 
If $\left(\Delta_1 c^o_{23}+\Delta_2 c^o_{31}+\Delta_3 c^o_{12}\right) = \underline{\lambda \neq 0}$, we use \eqref{e:e3} to solve for the constraint 
\begin{equation}\label{e:constraint}
	\kappa_V \left(\Delta_1 c^o_{23}+\Delta_2 c^o_{31}+\Delta_3 c^o_{12}\right) =
	\Delta_1 \left(c^o_2 c^e_3 - c^o_3 c^e_2\right) +
	\Delta_2 \left(c^o_3 c^e_1 - c^o_1 c^e_3\right) +
	\Delta_3 \left(c^o_1 c^e_2 - c^o_2 c^e_1\right)
	\,.
\end{equation}
When $\underline{\lambda=0}$, we need to ensure the RHS of equations \eqref{e:vev_sol} is also zero. 
Since $\lambda$ is the sum of these RHS's, for this case we must impose
\begin{equation}\label{e:lambda0}
	\begin{split}
		\Delta_2 c^o_3 - \Delta_3 c^o_2 = 0 \,,\\*[1mm]
		\Delta_3 c^o_1 - \Delta_1 c^o_3 = 0 \,,\\*[1mm]
		\Delta_1 c^o_2 - \Delta_2 c^o_1 = 0 \,,\\
	\end{split}
\end{equation}
and we still have the constraint \eqref{e:e3} between $\kappa_V$, $c^e_i$, and the vev components.

Equations \eqref{e:e1} and \eqref{e:e2} allow us to extract the rotation matrix entries $Q_{2i}$ from the couplings and the vev.
From these we can also get the line $R_{1i}$, using
\begin{equation}\label{e:RfromQ}
	\begin{split}
		&\begin{pmatrix}
			R_{11} & R_{12} & R_{13} 
		\end{pmatrix}
		\, = \,
		\begin{pmatrix}
			Q_{21} & Q_{22} & Q_{23} 
		\end{pmatrix}
		\,,\\*[1mm]
		&\begin{pmatrix}
			0 & R_{14} & R_{15}
		\end{pmatrix}
		\, = \, 
		\begin{pmatrix}
			Q_{24} & Q_{25} & Q_{26}
		\end{pmatrix}
		(R_H)^T
		\, .
	\end{split}
\end{equation}

Notice that,
in the limit where all the couplings are SM-like,
i.e. $\kappa_V = 1$, $c^e_i=1$, and $c^o_i=0$,
the constraints \eqref{e:e3}-\eqref{e:constraint} are all trivial,
and the vevs can be chosen freely.
There is, however, an interesting information to be garnered from
Eq.~\eqref{e:e4}.
Let us take $c_t^o=0$. Then, we find from \eqref{e:e4}
\begin{equation}
\frac{c_\tau^0}{c_b^0}
= - \frac{\hat{v}_2^2}{\hat{v}_1^2}
=  - \tan^2{\beta_1} \, ,
\label{anticorr}
\end{equation}
where we have used Eq.~\eqref{vev:param} on the last equality.
This equation means that,
perhaps unexpectedly because this is a type-Z model with a priori uncorrelated
charged lepton and down-quark couplings,
there is indeed a correlation among them.
Thus,
as we will see below,
one can expect that the current constraint on $c^o_\tau$ from
CMS \cite{CMS:2021sdq} and ATLAS \cite{ATLAS:2022akr}
will have some impact on $c^o_b$ as well.\footnote{Notice that equations (64)-(69), and, thus, the remaining analytical results in this section 3.3, are valid regardless of whether type-Z couplings are imposed by a $Z_2 \times Z_2$ symmetry or, alternatively, by a $Z_3$ symmetry. To that extent, our results are quite general. However, there will be numerical differences between the $Z_2 \times Z_2$ case discussed here and the $Z_3$ case. The reason is that both cases differ in the details of the couplings in the potential \cite{Boto:2023nyi}, which are limited by both theoretical and experimental constraints in different fashions. And, since the exact values for the couplings depend on the couplings in the potential, we could reach different numerical conclusions in both cases.}

\section{Constraints on the Parameter Space}
\label{ses:constraints}

In this section, we present the constraints that must be satisfied by the model parameters for theoretical and phenomenological consistency.

\subsection{Boundedness from below}

Necessary and sufficient BFB conditions have not yet been obtained for the $Z_2\times Z_2$ symmetric case.
However, sufficient conditions can be obtained in a way similar to \cite{Boto:2022uwv}, which consist in checking the copositivity of the matrix
\begin{equation}
	A = 
	\begin{pmatrix}
		\lambda_{11} & \hat{\lambda}_{12} & \hat{\lambda}_{13} \\
		\hat{\lambda}_{12} & \lambda_{22} & \hat{\lambda}_{23} \\
		\hat{\lambda}_{13} & \hat{\lambda}_{23} & \lambda_{33} 
	\end{pmatrix}
	\;,
\end{equation}
with
\begin{equation}
	\begin{split}
		\lambda_{11} &=  2\lambda_1  \;, \qquad
		\lambda_{22} =  2\lambda_2  \;, \qquad
		\lambda_{33} =  2\lambda_3  \;, \\
		\hat{\lambda}_{12} &= \lambda_{12} + \min(0,\lambda^\prime_{12}) - 2|\lambda_{10}|  \;, \\
		\hat{\lambda}_{13} &= \lambda_{13} + \min(0,\lambda^\prime_{13}) - 2|\lambda_{11}|  \;, \\
		\hat{\lambda}_{23} &= \lambda_{23} + \min(0,\lambda^\prime_{23}) - 2|\lambda_{12}|  \;.  \\
		\lambda_{12} &= \lambda_4 + \lambda_7 \;, \quad
		\lambda_{13} = \lambda_5 + \lambda_8 \;, \quad
		\lambda_{23} = \lambda_6 + \lambda_9 \;, \\
		\lambda_{12}^\prime &= - \lambda_7 \;, \qquad\;
		\lambda_{13}^\prime = - \lambda_8 \;, \qquad\;
		\lambda_{23}^\prime = - \lambda_9 \;. \\
	\end{split}
\end{equation}
The copositivity conditions are \cite{Klimenko:1984qx,Kannike:2012pe}.
\begin{equation}
	\begin{split}
		&A_{11} \geqslant 0 ,\; A_{22} \geqslant 0 ,\; A_{33} \geqslant 0  \;,\\
		&\overline{A_{12}} = \sqrt{A_{11}A_{22}} + A_{12} \geqslant 0 ,\;
		\overline{A_{13}} = \sqrt{A_{11}A_{33}} + A_{13} \geqslant 0 ,\;
		\overline{A_{23}} = \sqrt{A_{22}A_{33}} + A_{23} \geqslant 0 ,\;
		\\
		&\sqrt{A_{11}A_{22}A_{33}} + A_{12}\sqrt{A_{33}} + A_{13}\sqrt{A_{22}} + A_{23}\sqrt{A_{11}} + \sqrt{2\overline{A_{12}}\;\overline{A_{13}}\;\overline{A_{23}}} \geqslant 0 \;.
	\end{split}
\end{equation}

\subsection{Perturbativity}

For the theoretical constraints, we first ensure the perturbativity of the
Yukawa couplings. For the type-Z Yukawa structure,
the top, bottom, and tau Yukawa couplings are given by 
\begin{eqnarray}\label{eq:yukawa}
	y_t = \frac{\sqrt{2}\, m_t }{v \sin\beta_2}\;,
	\quad y_b = \frac{\sqrt{2}\, m_b }{v \sin\beta_1 \cos\beta_2} \;,
	\quad y_\tau = \frac{\sqrt{2}\, m_\tau }{v \cos\beta_1 \cos\beta_2} \;,
\end{eqnarray}
which follow from our convention that $\Phi_3$, $\Phi_2$, and $\Phi_1$
couple to up-type quarks, down-type quarks, and charged leptons respectively.
To maintain the perturbativity of Yukawa couplings, we impose
$\lvert y_t \rvert,\lvert y_b\rvert,\lvert y_\tau\rvert < \sqrt{4\pi}$. 
Throughout our paper, we have used values of $\tan\beta_{1,2}$
which are consistent with this perturbative region.

\subsection{Unitarity}

The unitarity bounds for our model have been obtained in \cite{Bento:2022vsb}.
These bounds correspond to imposing that the absolute value of the eigenvalues of
the scattering matrices is smaller than $8\pi$. In section 5.10 of
\cite{Bento:2022vsb} these matrices are written using the 3HDM
notation of \cite{Ferreira:2008zy},
\begin{equation}
	\begin{split}
		r_i = \lambda_i \;,&\quad i=1,2,3 \;,\\
		r_i = \frac{1}{2} \lambda_i \;,&\quad i=4,5,6,7,8,9 \;,\\
		c_3 =  \lambda_{10}  \;, \qquad
		c_5 &=  \lambda_{11} \;, \qquad
		c_{17} =  \lambda_{12}  \;, 
	\end{split}
\end{equation}
\begin{equation}
	M_2^{++} =  2\; \text{diag}\left\{
	\begin{pmatrix}
		r_1 & c_3 & c_5 \\
		c_3^* & r_2 & c_{17} \\
		c_5^* & c_{17}^* & r_3
	\end{pmatrix},
	(r_4+r_7),
	(r_5+r_8),
	(r_6+r_9)
	\right\} \;,
\end{equation}
\begin{equation}
	M_2^{+} =  2\; \text{diag}\left\{
	\begin{pmatrix}
		r_1 & c_3 & c_5 \\
		c_3^* & r_2 & c_{17} \\
		c_5^* & c_{17}^* & r_3
	\end{pmatrix},
	\begin{pmatrix}
		r_4 & r_7 \\
		r_7 & r_4
	\end{pmatrix},
	\begin{pmatrix}
		r_5 & r_8 \\
		r_8 & r_5
	\end{pmatrix},
	\begin{pmatrix}
		r_6 & r_9 \\
		r_9 & r_6
	\end{pmatrix}
	\right\} \;,
\end{equation}
\begin{equation}
	M_0^{+} =  2\; \text{diag}\left\{
	\begin{pmatrix}
		r_1 & r_7 & r_8 \\
		r_7 & r_2 & r_9 \\
		r_8 & r_9 & r_3
	\end{pmatrix},
	\begin{pmatrix}
		r_4 & c_3 \\
		c_3^* & r_4
	\end{pmatrix},
	\begin{pmatrix}
		r_5 & c_5 \\
		c_5^* & r_5
	\end{pmatrix},
	\begin{pmatrix}
		r_6 & c_{17} \\
		c_{17}^* & r_6
	\end{pmatrix}
	\right\} \;,
\end{equation}
\begin{equation}
	M_2^0 = M_2^{++} \;,
\end{equation}
\begin{equation}
	\begin{split}
		M_0^{0} \sim  2\; \text{diag}\Bigg\{
		\frac{1}{2}M_0^+,&
		\begin{pmatrix}
			3r_1 & 2r_4+r_7 & 2 r_5+r_8 \\
			2r_4+r_7 & 3r_2 & 2r_6+r_9 \\
			2 r_5+r_8 & 2r_6+r_9 & 3r_3
		\end{pmatrix},
		\begin{pmatrix}
			r_4+2r_7 & 3c_3 \\
			3c_3^* & r_4+2r_7
		\end{pmatrix},\\&
		\begin{pmatrix}
			r_5+2r_8 & 3c_5 \\
			3c_5^* & r_5+2r_8
		\end{pmatrix},
		\begin{pmatrix}
			r_6+2r_9 & 3c_{17} \\
			3c_{17}^* & r_6+2r_9
		\end{pmatrix}
		\Bigg\} \;,
	\end{split}
\end{equation}
where the equality $=$ of scattering matrices is valid up to permutations,
and they are similar $\sim$ if they are equal up to an orthogonal basis change.
It was also shown in \cite{Bento:2022vsb} that, instead of computing the
eigenvalues of the $3\times3$ matrices, it is faster to check that the
eigenvalues are bounded using the equivalence between the following two statements,
regarding the eigenvalues $\lambda_i$ of an $n\times n$ hermitian matrix $A$:

\begin{enumerate}
	\item the eigenvalues are bounded by $|\lambda_i|<c$;
	\item  $D_k(A+c\one)>0$ and $D_k(-A+c\one)>0$, where $D_k$ is the determinant of the upper left $k\times k$ submatrix.
\end{enumerate}

\subsection{Oblique Parameters STU}

We use the results in \cite{Grimus:2007if}, using matrices $U$ and $V$, defined in equations \eqref{matrixU} and \eqref{matrixV}, and the fit results in \cite{Baak:2014ora}.

\subsection{Other Constraints}

Besides the constraints described previously, BFB,
Unitarity and STU, we have to consider all other theoretical and
experimental constraints. From these, the most important are those
coming from the LHC experiments at CERN. First, we must make sure that
we comply with some experimental features of a SM-like Higgs.
For this we compute the Higgs signal
strengths $\mu^f_i$ defined by,
\begin{eqnarray}
	\mu_i^f =\left(\frac{\sigma_i^{\text{3HDM}}(pp\to h_{125})
  }{\sigma_i^{\text{SM}}(pp\to
  h_{125})}\right)\left(\frac{\text{BR}^{\text{3HDM}}(h_{125}\to
  f)}{\text{BR}^{\text{SM}}(h_{125}\to f)}\right) \,, 
	\label{e:ss}
\end{eqnarray}
where the subscript `$i$' denotes the production mode,
and the superscript `$f$'
denotes the decay channel of the SM-like Higgs scalar.
Starting from the collision of two protons, the relevant production
mechanisms include gluon fusion~($ggF$), vector boson fusion~($VBF$),
associated production with a vector boson ($VH$, $V = W$ or $Z$), and
associated production with a pair of top quarks ($ttH$).
For the $ggF$ we use HIGLU~\cite{Spira:1995mt} at NNLO. The
effect of the bottom quark is included and it is more important as
$\tan\beta$ increases. The effect of the interference between top and
bottom quark loops is also included by calculating in the SM with only
top, only bottom and both, and from there infer the
interference. After this, the SM results are multiplied by the couplings
in the model to reconstruct the total cross section\footnote{We have
compared with the cross sections in \texttt{HiggsTools} and found an
agreement within a few percent.}. For the $VBF$, $VH$ and $ttH$,
we just want the ratios of the model cross section with respect to the
SM -- see Eq.~\eqref{e:ss} -- and these are given by $\kappa_V^2$ for
the first two~\cite{Fontes:2014xva,deFlorian:2016spz}.  For $ttH$,
the pure CP even and the pure CP odd cross sections are
different~\cite{Frederix:2011zi,Broggio:2017oyu}, so we multiply the
SM result by,
\begin{equation}
  \label{eq:7}
(c_t^e)^2+  \frac{\sigma^{\rm ttH}_{0^-}}{\sigma^{\rm ttH}_{0^+}}\
(c_t^o)^2  \, ,
\end{equation}
for this case. For $M_H=125$ GeV and $\sqrt{s}=13$ TeV, we get from
Ref.\cite{Broggio:2017oyu} that $\sigma^{\rm
  ttH}_{0^-}/\sigma^{\rm ttH}_{0^+} \simeq 0.416$, at NLO+NLL
order.\footnote{We see that for
    \texttt{ttH}, the pure pseudoscalar 
  cross section is smaller 
than the pure scalar one, in contrast with gluon-gluon fusion where
the pure pseudoscalar cross section is larger than the scalar one by a
factor of 9/4~\cite{Jaquier:2019bfs}.}
Because $(c_t^o)^2$ is very small (see
section~\ref{sec:results} below) the factor in Eq.~(\ref{eq:7}) is
always very close to $(c_t^e)^2$.
The SM cross sections for these production modes are obtained from
\url{https://twiki.cern.ch/twiki/bin/view/LHCPhysics/LHCHWGCrossSectionsFigures}.
Our code calculates all the decays of the Higgs boson $h_{125}$, using the
Feynman rules derived with
\texttt{FeynMaster}\cite{Fontes:2019wqh,Fontes:2021iue}.
In particular
for the important channel $h\to \gamma\gamma$ we adapt our previous
results \cite{Fontes:2014xva}, where a full description of the channel
$ h\to \gamma Z$ is also given in detail.
We demand that the
predicted signal rates agree within $2 \sigma$ with each individual
signal rate measurement, as given by the ATLAS results summarized in
Fig.~3 of Ref.~\cite{ATLAS:2022vkf}.  The ATLAS measurements are well
in agreement with the corresponding CMS results \cite{CMS:2022dwd},
such that all our conclusions would remain unchanged if, instead, the
CMS results or a combination of ATLAS+CMS results were used. 

For comparison, we also will consider an alternative method to 
incorporate the various signal strengths,
by using the HiggsSignals module inside  \texttt{HiggsTools-1.1.3} \cite{Bahl:2022igd}.
This method has the advantage that it takes into account correlations
amongst different observables.
However, as we will show, 
the conclusions are largely insensitive to the difference between approaches.

Next we must make sure that we comply with the LHC results on the
direct searches for other scalars, either neutral or charged. For this
we use the software package \texttt{HiggsTools-1.1.3}
\cite{Bahl:2022igd} which includes the latest data from the ATLAS and
CMS experiments at CERN. Our code calculates all the needed couplings
and decay widths for those nonstandard scalars, allowing for decays
with off-shell scalar bosons, using the method explained in
Ref.~\cite{Romao:1998sr}.

The next step is to take into consideration the bounds coming from
flavour data.
In the type-Z 3HDM there are no FCNCs at the tree-level. Therefore the
NP contribution at 
one-loop order to observables such as $b\to s\gamma$ and the neutral meson mass
differences will come from the charged scalar Yukawa couplings.
It was found in
Ref.~\cite{Chakraborti:2021bpy} that the constraints coming from the meson mass
differences tend to exclude very low values of $\tan\beta_{1,2}$. Therefore, we
only consider
\begin{eqnarray}
	\tan\beta_{1,2} > 0.3 \,,
\end{eqnarray}
to safeguard ourselves from the constraints coming from the neutral meson mass
differences. To deal with the constraints stemming from  $b\to s\gamma$, we
follow the procedure described in Refs.~\cite{Florentino:2021ybj,Boto:2021qgu, Akeroyd:2020nfj}
and impose the following restriction
\begin{equation}
	\label{e:b2sg}
	2.87 \times 10^{-4} < \text{BR}(B\to X_s \gamma) < 3.77 \times 10^{-4}\,,
\end{equation}
which represents the $3\sigma$ experimental limit.

Finally, in addition to these, as our model is no longer explicitly CP
conserving, 
we can now have, in principle, a non-zero electric dipole moment of the electron
(eEDM), which is highly constrained experimentally
\cite{ACME:2013pal,ACME:2018yjb,Roussy:2022cmp}.
For the theoretical predictions of the eEDM, we use the formulae in
\cite{Barr:1990vd,Yamanaka:2013pfn,Abe:2013qla,Inoue:2014nva,Altmannshofer:2020shb}.

\section{How we performed the scan}
\label{sec:scan}

In order to perform our scan we may choose a specific way to parameterize
$R$ in Eq.~\eqref{mass_eigenstates} as some product of the ten rotations $R_{ij}$ in five dimensions.
$R_{ij}$ are $5 \times 5$ matrices which are like the identity matrix in all entries
except the entries $ii$ and $jj$, given by $\cos{\alpha_{ij}}$,
the entry $ij$, given by $\sin{\alpha_{ij}}$,
and the entry $ji$ given by $-\sin{\alpha_{ij}}$.
To be specific, for our scans we have chosen
\begin{eqnarray}
R &=& R_{CPV} R_x R_z = R_{CPV} R_z R_x \;,
\nonumber\\
R_x &=& R_{23} R_{13}R_{12} \;,
\nonumber\\
R_z &=& R_{45} \;,
\nonumber\\
R_{CPV} &=& R_{35}R_{34}R_{25}R_{24}R_{15}R_{14}\, .
\label{eq:Rspecific}
\end{eqnarray}

\subsection{The random scan}
Ideally one would like to perform
an extensive scan of the parameter space.
Our fixed inputs are $v = 246\,\text{GeV}$ and $m_{h1} = 125\,\text{GeV}$.
We then would take random values in the
ranges:
\begin{align}
  &\theta,\,\varphi,\,\alpha_{12},\, \alpha_{13},\, \alpha_{14},\,
  \alpha_{15},\, \alpha_{23},\, \alpha_{24},\,\alpha_{25},\,
  \alpha_{34},\, \alpha_{35},\,\alpha_{45}\, \in
\left[-\frac{\pi}{2},\frac{\pi}{2}\right];\\[8pt]
&\tan{\beta_1},\,\tan{\beta_2}\,\in \left[0.3,10\right]; 
\\[8pt]
& m_{h_2},\, 
\in \left[125,1000\right]\,\text{GeV},\,
m_{H_1^\pm},\,m_{H_2^\pm}\,
\in \left[100,1000\right]\,\text{GeV};\\[8pt]
&
\text{Re}(m^2_{12}),\text{Re}(m^2_{13}),
\text{Re}(m^2_{23}) \in  \left[\pm 10^{-1},\pm 10^{7}\right]\,
\text{GeV}^2\, .
\label{eq:scanparameters}
\end{align}
These 20 free parameters (other than $v$ and $m_{h_1}$)
fully determine the point in parameter space,
as we had found from the Lagrangian in section~\ref{subsec:paramphys}.
As explained before, the masses $m_{h_3},\, m_{h_4},\, m_{h_5}$, and the
remaining parameters of the potential are obtained from these 20
parameters (plus $v$ and $m_{h_1}$).
However, as in the real 3HDM case, this random scan has a very low
probability of success\cite{Boto:2022uwv,Boto:2021qgu,Das:2019yad}. In
fact even worse than in the real case; 
lower than 1 point per $10^{13}$ sampled
points\footnote{ We sampled $10^{13}$ points using
    2100 CPU hours (on an AMD Ryzen 9 7950X 16-Core processor) and did not find any good point.}.

\subsection{Importing from the real 3HDM}

We had therefore to follow another strategy to have points.
As we have seen in section~\ref{subsec:real},
the C3HDM has a real limit corresponding to,
\begin{align}
  \label{eq:1}
  &\alpha_{12}=\alpha_{1},\ \ 
  \alpha_{13}=\alpha_{2},\ \ \alpha_{23}=\alpha_{3},\ \ 
  \alpha_{45}=-\gamma_1,\ \ \theta=-\gamma_2 \nonumber\\[8pt]
  &\varphi=\alpha_{14}=\alpha_{15} = \alpha_{24}=\alpha_{25}=
  \alpha_{34}=\alpha_{35}=0\, ,
\end{align}
where $\alpha_1,\,\alpha_2,\,\alpha_3,\,\gamma_1,\,\gamma_2$ are the
variables defined in the real 3HDM\cite{Boto:2021qgu,Boto:2022uwv},
and $\beta_1,\,\beta_2$ retain their meaning in both cases.
So we produced sets of points in the real 3HDM, following a strategy
of closeness to the alignment limit as set out in \cite{Das:2019yad,Boto:2023nyi}.
And then we imported them into
the C3HDM with the assignments in Eq.~(\ref{eq:1}). These points
should also be good C3HDM points. The next step is to scan around
these points. This was done within the ranges
\begin{align}
  \label{eq:2}
  \alpha_{14},&\, \alpha_{15} \in [-0.01,0.01];\nonumber\\[8pt]
  \varphi,\,\alpha_{24},\,&\alpha_{25},\,\alpha_{34},\,\alpha_{35} \in
  [-0.1,0.1] .
\end{align}
These ranges were chosen to be close to the real case. We find out
that  $\alpha_{14},\, \alpha_{15}$ are more directly connected with
the eEDM and therefore have to be smaller in our scans to comply with
that limit.

\begin{figure}[htb]
  \centering
  \begin{tabular}{cc}
      \includegraphics[width=0.48\textwidth]{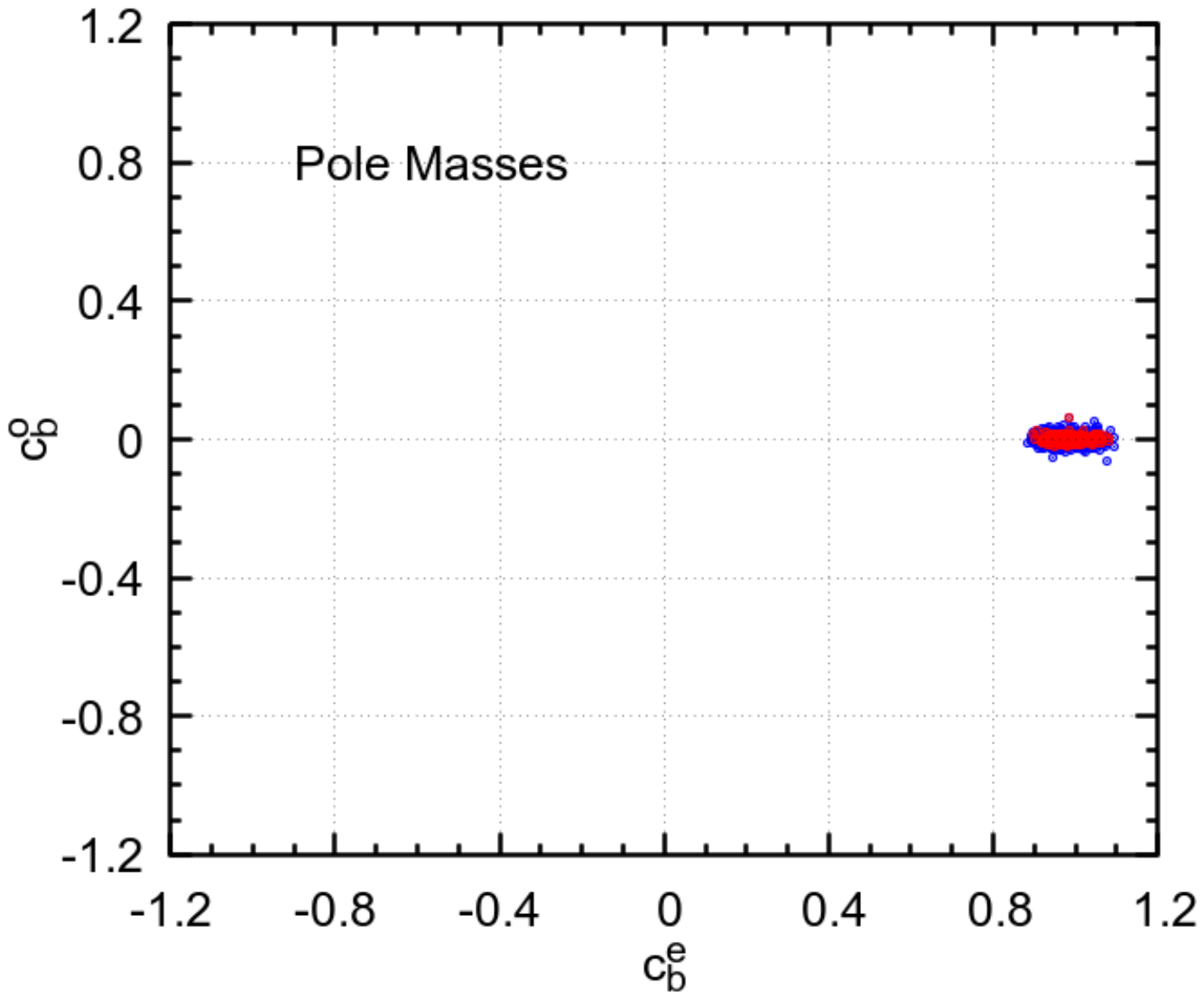}
    &
    \includegraphics[width=0.48\textwidth]{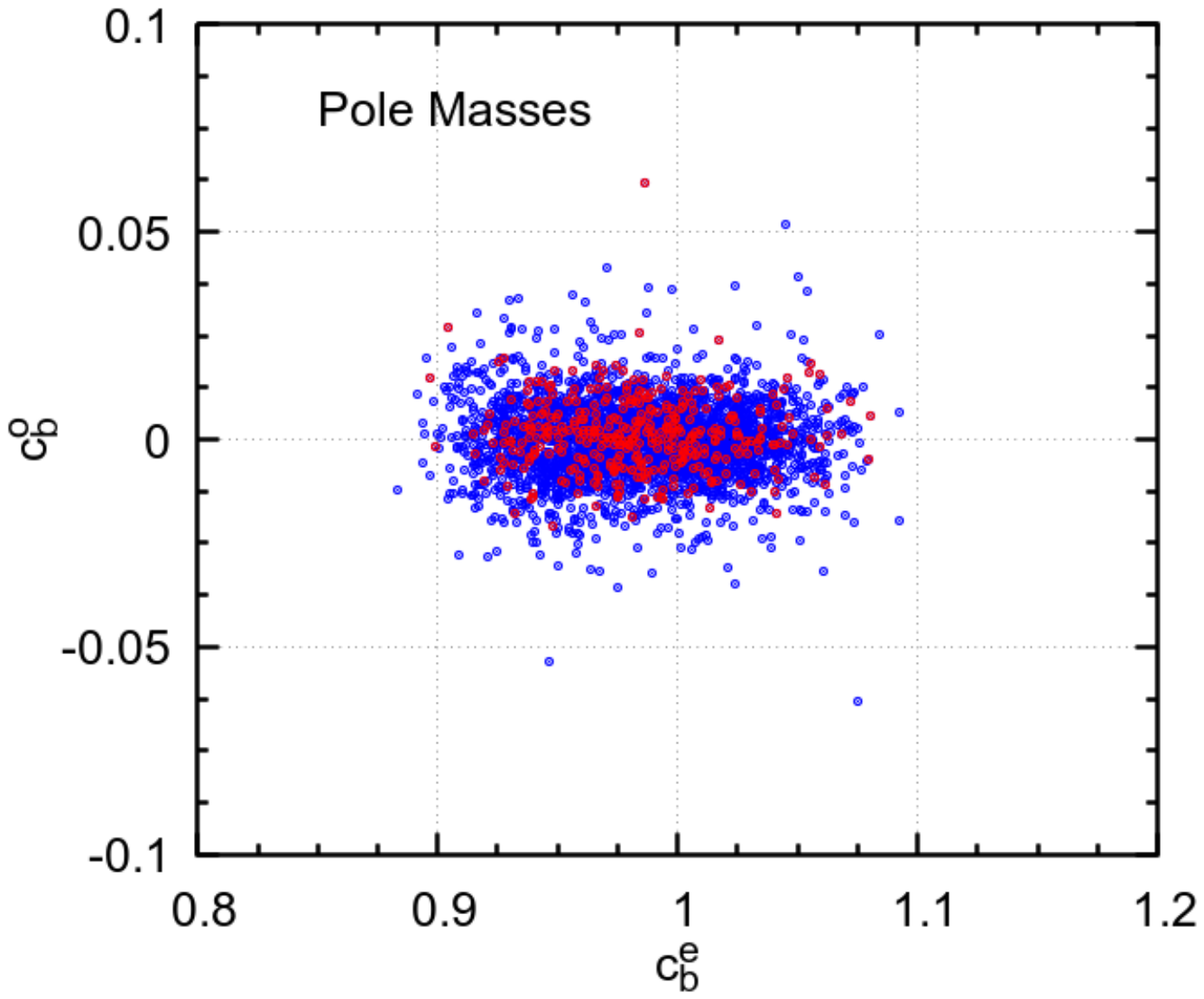}
  \end{tabular}
  \caption{Points obtained by scanning around solutions of the real
    3HDM. The blue (red) points are before (after) the
    \texttt{\rm HiggsTools} constraints.} 
  \label{fig:3}
\end{figure}

In Fig.~\ref{fig:3} we show the result of this procedure. On the
left panel we plot in the whole expected
region\cite{Fontes:2017zfn,Biekotter:2024ykp}, while in the right
panel we have a close up. These points are for pole masses for the
eEDM calculation, but a similar result would be true for the
parameters at scale $M_Z$. For a discussion of why this is important
see Ref.\cite{Biekotter:2024ykp}.
We remark that this is a very inefficient way of generating good C3HDM
points, although it seems the only
one\footnote{\label{foot:AI}We intend to apply the techniques developed in
  Refs.\cite{deSouza:2022uhk,Romao:2024gjx} to explore fully the C3HDM
  parameter space. In fact in the present way we are not sure of
  covering all regions of the parameter space away from the real
  3HDM.}. We started from 81386 points 
from the real 3HDM, already after passing the
\texttt{HiggsTools}\cite{Bahl:2022igd} constraints. From these, only
2893 passed all the C3HDM constraints (in blue in Fig.~\ref{fig:3})
and after \texttt{HiggsTools} only 372 remained (in red). This so
large difference is easy to understand. Once we generate the C3HDM
parameters, the masses $m_{h_3},\, m_{h_4},\, m_{h_5}$, are derived
quantities, which have to satisfy being in the interval $[125,1000]$
GeV and the STU constraints (these could have been true for the real 3HDM
points, but we have no control over these masses when we import the
points into the C3HDM). But most importantly, the calculation of the
parameters of the 
potential also changes and many points are discarded due to BFB or
perturbative unitarity constraints.

\subsection{Enlarging the pseudoscalar component}

\begin{figure}[htb]
  \centering
  \begin{tabular}{cc}
      \includegraphics[width=0.48\textwidth]{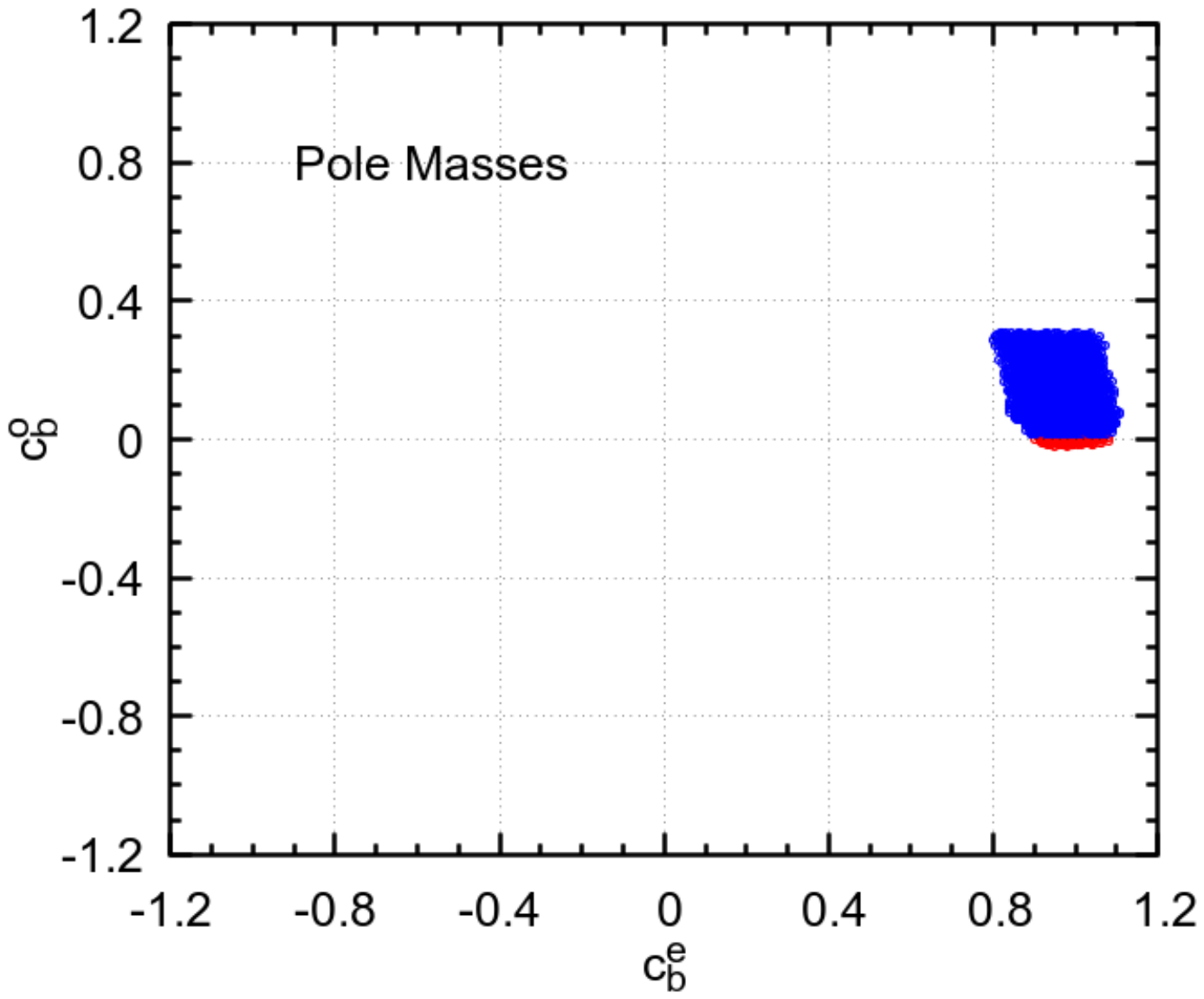}
    &
    \includegraphics[width=0.48\textwidth]{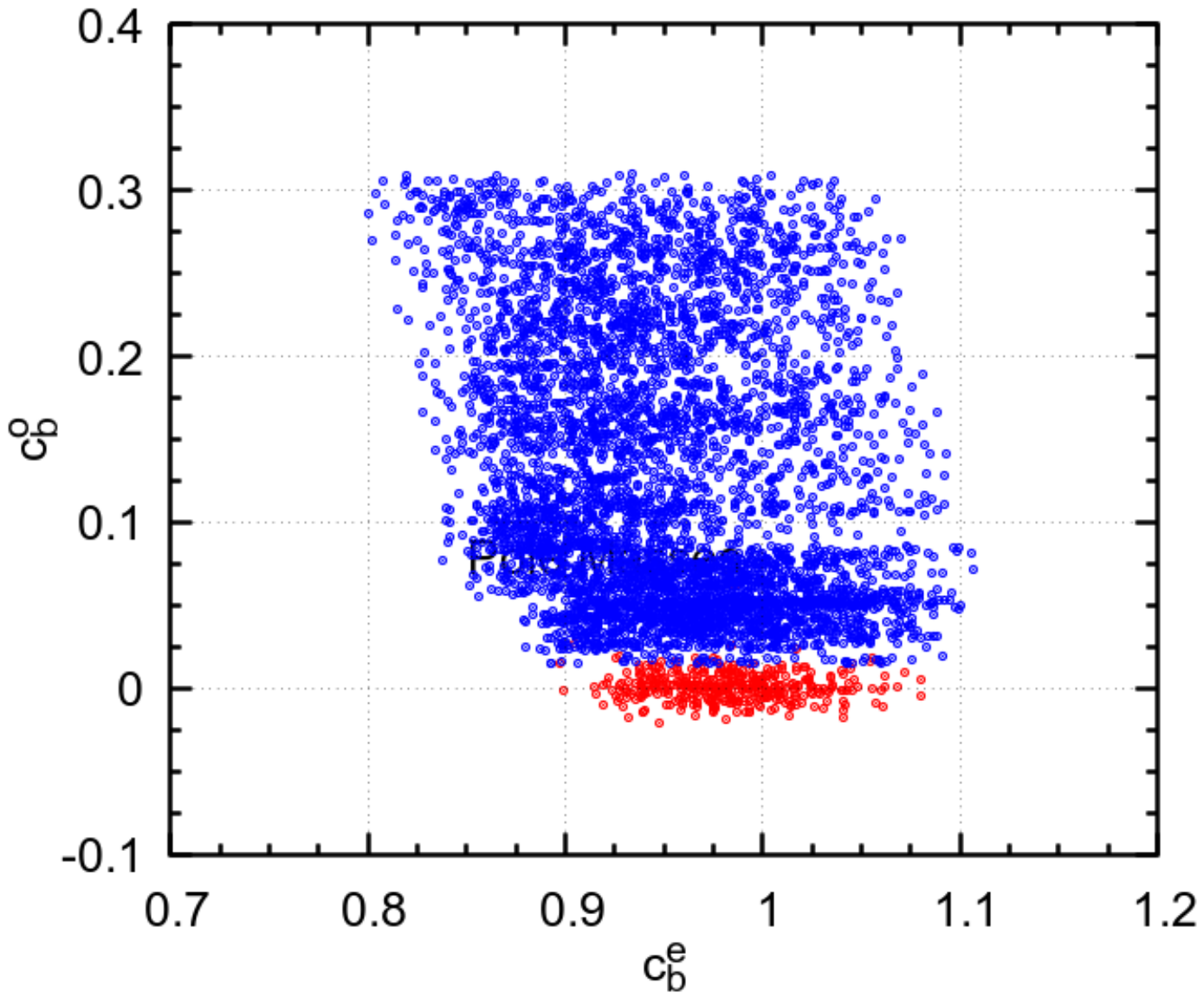}
  \end{tabular}
  \caption{Enlarging the pseudoscalar component. In the left panel we
    show the whole
    range, while in the right panel we show a closeup. The blue points
  were obtained from the red points as explained in the text.}
  \label{fig:4}
\end{figure}

The points shown in Fig.~\ref{fig:3} have non zero pseudoscalar
components, but those are small because of the way in which they were generated,
coming from the real 3HDM case. So the question arises: can one
enlarge the pseudoscalar component? To answer this question we
developed the following strategy. We start with the good C3HDM set of points
after \texttt{HiggsTools}, in red in Fig.~\ref{fig:3}. We then
consider random points \textit{around} those points. From this run, we select
only the points with the largest pseudoscalar component (supposing,
for the moment that we are going in the positive direction, the same
can be applied going into the negative direction, of course). We
consider these as input and scan around these points, keeping only the
points with the largest pseudoscalar component. We repeat this
procedure until we cannot go any higher. Much of the success depends,
clearly, on the meaning of \textit{around}, and of course, this is a
time consuming method. We present our results in
section~\ref{sec:results}, after discussing the wrong sign.
But,
just for illustration,
we present in Fig.~\ref{fig:4} the situation after a few
iterations.
The points in red are those of Fig.~\ref{fig:3}, generated from the
initial points from the real 3HDM. Points in blue were generated by
the method explained above\footnote{As mentioned above, this is a time
  consuming method because it is difficult to automatize.}.

\subsection{The wrong sign}

\begin{figure}[htb]
  \centering
  \begin{tabular}{cc}
      \includegraphics[width=0.48\textwidth]{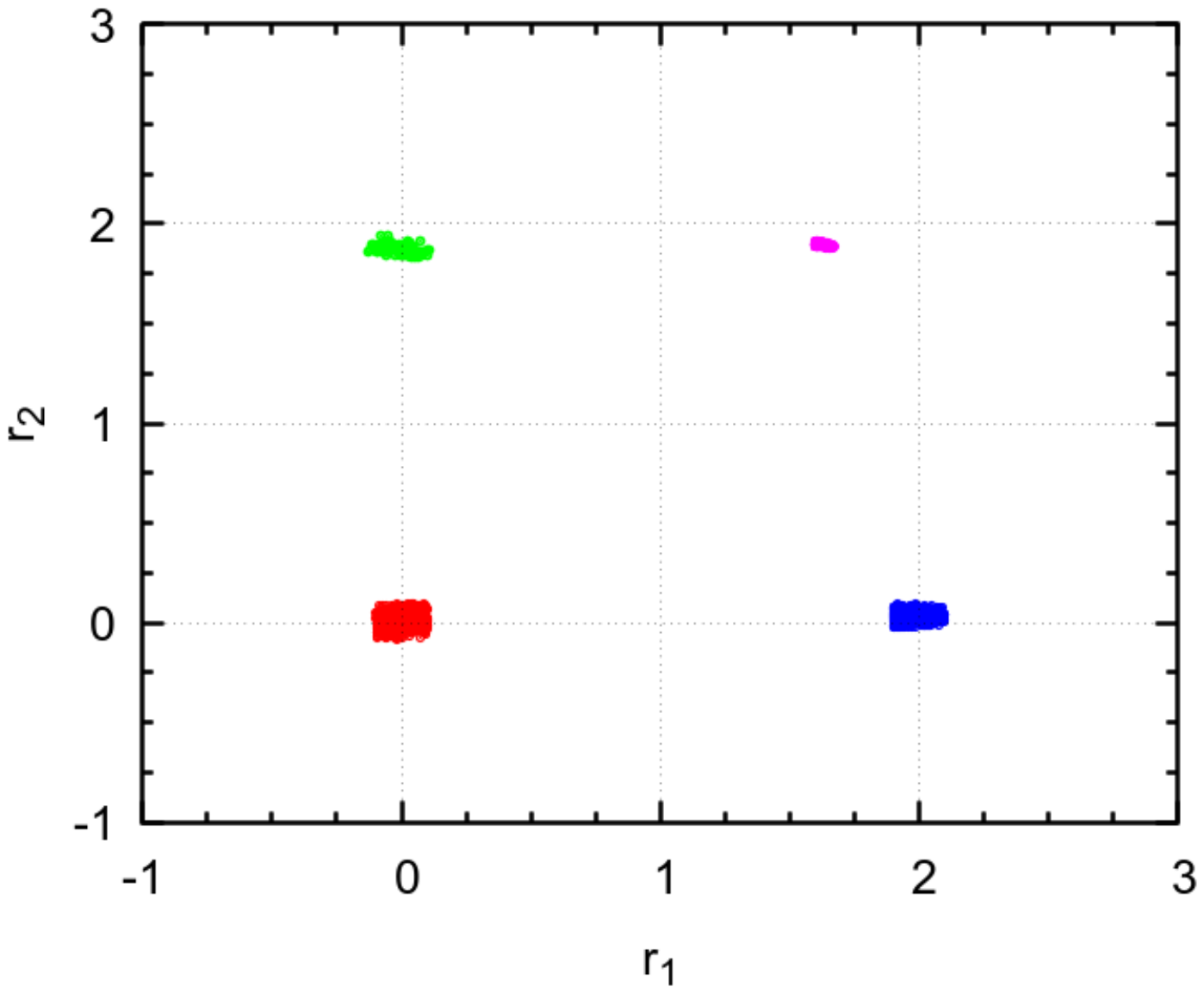}
    &
    \includegraphics[width=0.48\textwidth]{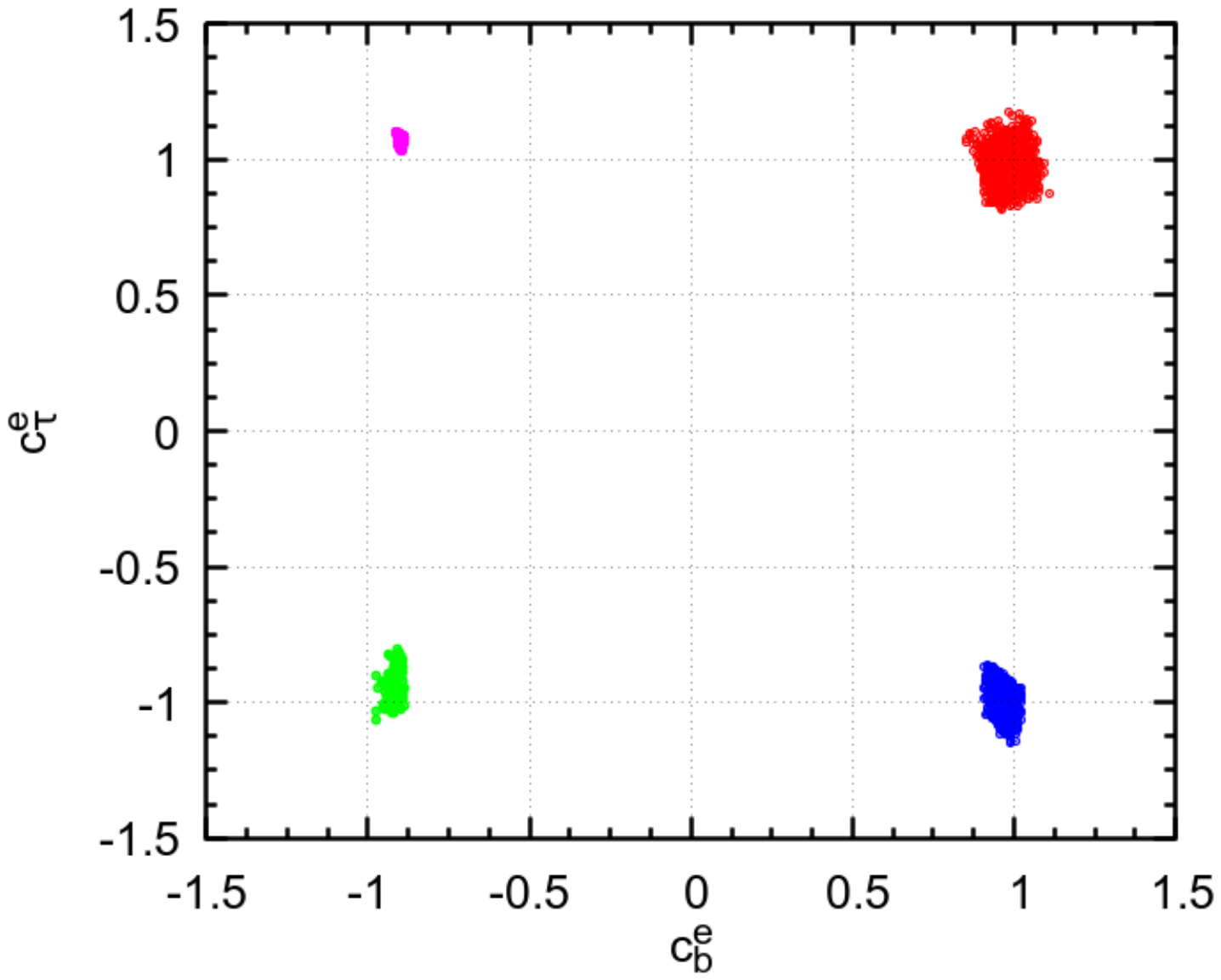}
  \end{tabular}
  \caption{In the left panel we plot the variables $r_i$, defined in
    Eq.~(\ref{eq:5}). In the right panel we show the corresponding
    values for the Yukawa couplings of the bottom quark and $\tau$
    lepton. The colors match in the two panels.}
  \label{fig:5}
\end{figure}

Up to now we have been discussing points that are not far away from
the \textit{correct} sign,
meaning,
\begin{equation}
  \label{eq:3}
  \kappa_V\simeq\kappa_U\simeq\kappa_D\simeq\kappa_L\simeq 1 ,
\end{equation}
where for type-Z in the
3HDM\cite{Boto:2021qgu,Boto:2022uwv,Das:2019yad,Das:2022gbm},  
\begin{align}
  \label{eq:4}
  &\kappa_V=\cos(\alpha_2) \cos(\alpha_1-\beta_1) \cos(\beta_2) +
  \sin(\alpha_2) \sin(\beta_2)\, ,\\[8pt]
  &  \kappa_U= \frac{\sin(\alpha_2)}{\sin(\beta_2)},
\quad
  \kappa_D=\frac{\sin(\alpha_1) \cos(\alpha_2)}{\sin(\beta_1) \cos(\beta_2)},
\quad
  \kappa_L= \frac{\cos(\alpha_1) \cos(\alpha_2)}{\cos(\beta_1) \cos(\beta_2)}.
\end{align}
It  was shown by Das and collaborators\cite{Das:2022gbm} that the type
Z in the real 3HDM offers more flexibility than the
C2HDM\cite{Fontes:2017zfn,Biekotter:2024ykp} with
respect to the possibility of the wrong sign. In fact they have shown
that while keeping $\kappa_V\simeq\kappa_U\simeq 1$ we can have
$\kappa_{D,L}\simeq \pm 1$. For this to happen they identified two
variables
\begin{equation}
  \label{eq:5}
  r_1=\sin(\alpha_1-\beta_1) \tan(\beta_1),\quad
  r_2=\sin(\alpha_2-\beta_2) \tan(\beta_2),
\end{equation}
that in the limit of  $\kappa_V\simeq\kappa_U\simeq 1$ give,
\begin{equation}
  \label{eq:6}
  \kappa_D\simeq 1-r_2,\quad \kappa_L\simeq (1-r_2)(1-r_1).
\end{equation}
The values $r_{1,2}=0,2$ are, thus, special.
\begin{table}[htb]
  \centering
  \begin{tabular}{|c|c|c|}\hline
    &$r_1=0$& $r_1=2$\\\hline
    $r_2=0$ &$\kappa_D=1\ \kappa_L=1$ & $\kappa_D=1\ \kappa_L=-1$ \\[+4pt]\hline
    $r_2=2$ &$\kappa_D=-1\ \kappa_L=-1$  & $\kappa_D=-1\ \kappa_L=1$
    \\[+4pt]\hline 
  \end{tabular}
  \caption{The wrong sign possibilities in the real 3HDM\cite{Das:2022gbm}.}
  \label{tab:wronsign}
\end{table}
Therefore, to study this possibility within the C3HDM, we started by
generating points in the real 3HDM that satisfy the conditions in
Table~\ref{tab:wronsign}. The result is shown in Fig.~\ref{fig:5}.
On the left panel we show the selected regions in the $r_1, r_2$
plane. All these points passed all the constrains, including the
bounds from LHC using \texttt{HiggsTools}. The points in magenta show
the closest we could get to the $r_1=r_2=2$ case. Nevertheless the
situation described in Table~\ref{tab:wronsign} is clearly shown in
the right panel. We have three distinct regions for the wrong sign,
and one (in red) for the normal sign situation that we studied before.

Using these wrong sign regions as input in the C3HDM, we were able to
generate points in the C3HDM for the cases $r_1=2,r_2=0$
($c^e_b=1,c^e_\tau=-1$) and for $r_1=2,r_2=2$ ($c^e_b=-1,
c^e_\tau=1$). For the case 
$r_1=0,r_2=2$, our method did not generate any good points of the
C3HDM, after all the constraints were taken in account.
However,
all these points have very small pseudoscalar components,
of the order of $10^{-2}$ or smaller.
Our method of enlarging the pseudoscalar component did not
work here. As we said before, this is not a proof that such components
do not exist, just that our method of starting from the real does not
work. Other methods should be used, as already referred to in
footnote~\ref{foot:AI}.

\section{Results}
\label{sec:results}

\begin{figure}[h!]
  \centering
  \begin{tabular}{cc}
    \includegraphics[width=0.48\textwidth]{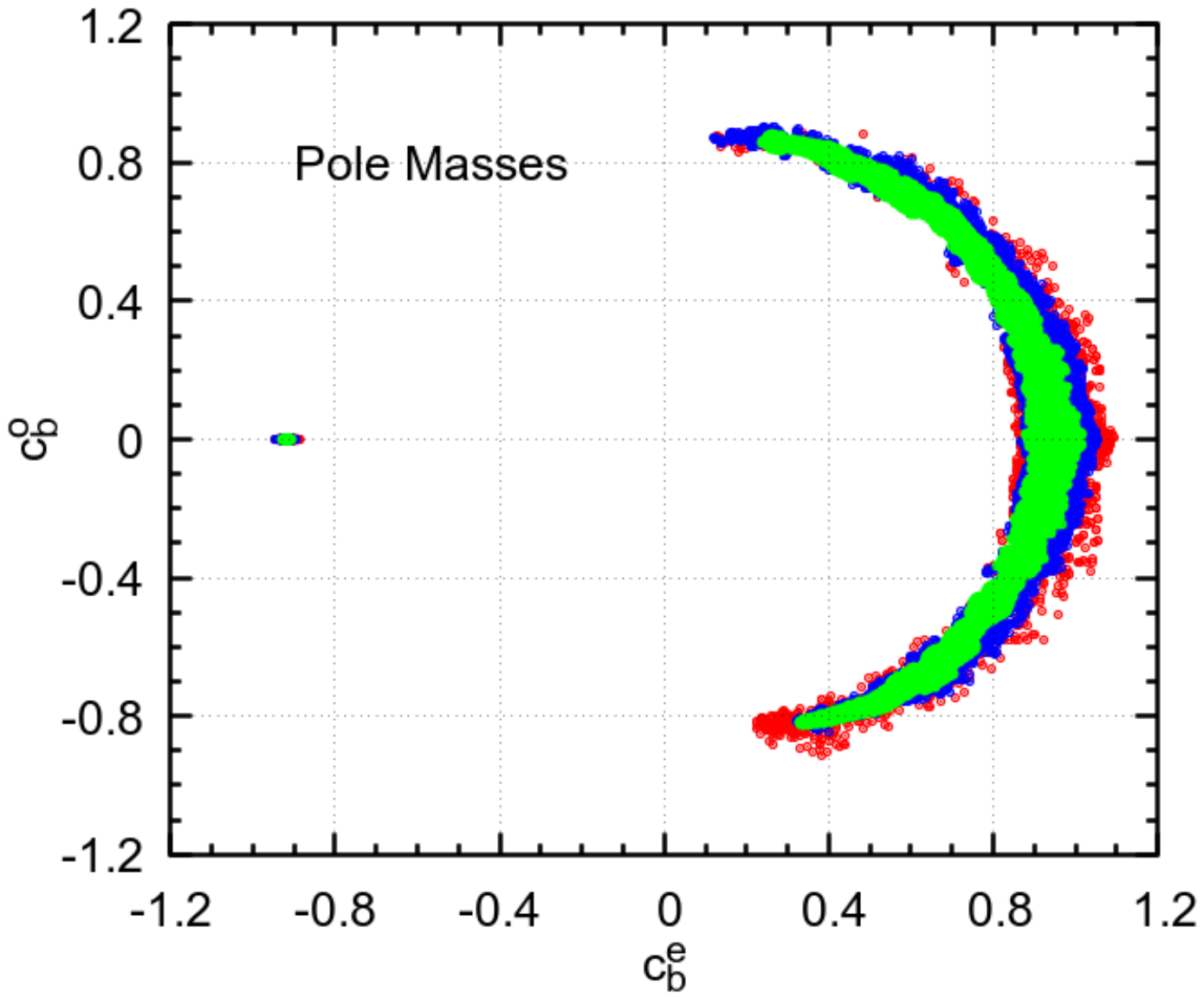}
    &
    \includegraphics[width=0.48\textwidth]{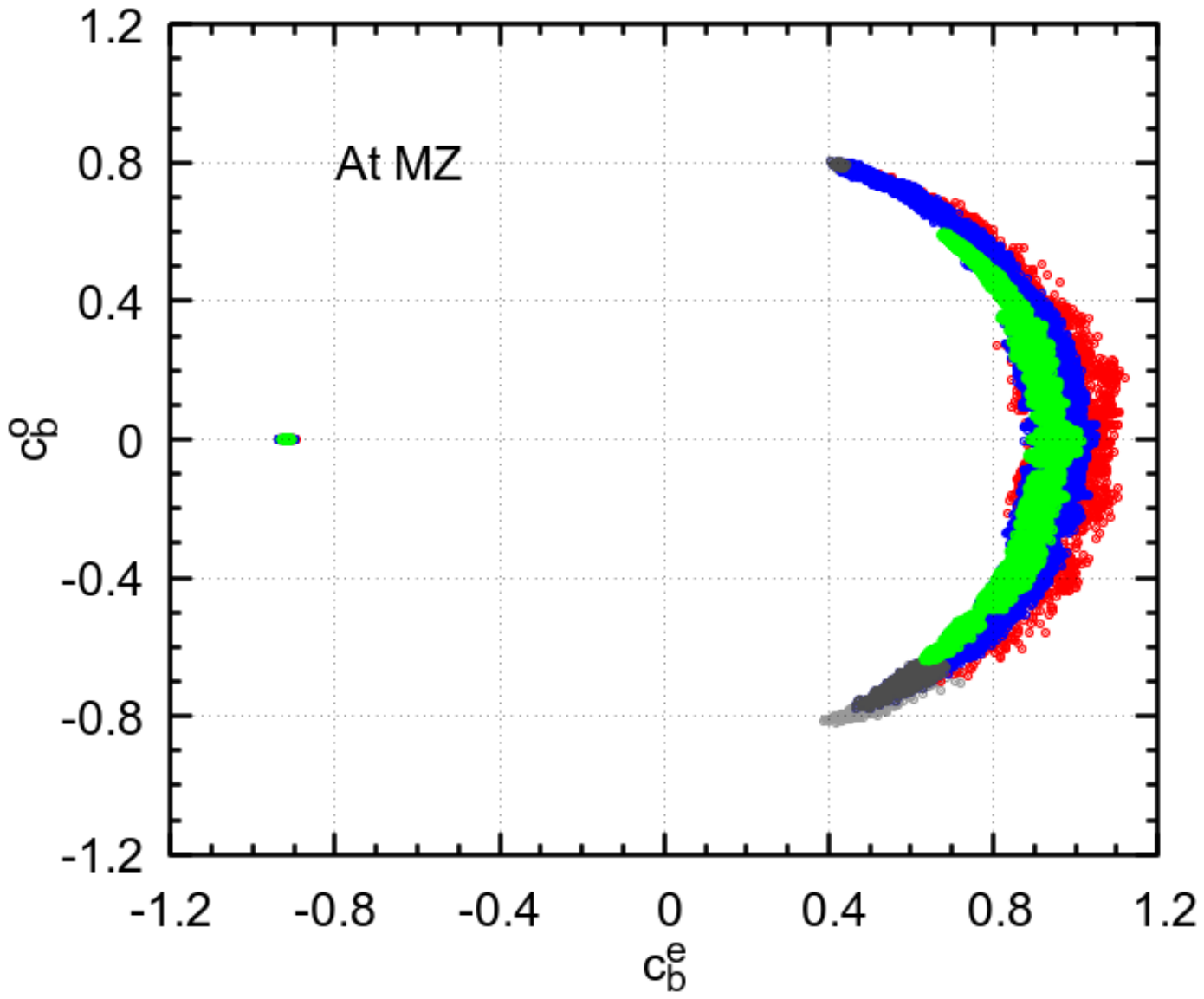}
  \end{tabular}
  \caption{In the left panel we plot $c^o_b$ vs. $c^e_b$  for the pole
    masses and right panel for 
    masses at the $M_Z$ scale. The constraints on $h_{125}\to
    \tau\bar{\tau}$ do not affect the former but slightly affect the
    later. See text for color code and Fig.~\ref{fig:1b} for more details. }
  \label{fig:1}
\end{figure}

In this section we present the combined results of the previous
scans.
In Fig.~\ref{fig:1}  
we present the results for the scalar and
pseudoscalar coupling of the down quarks.
The
points in red correspond to a $2 \sigma$ agreement with each individual
signal strength measurement.
The points in blue and green correspond to the method where
the signal strengths are constrained using  HiggsSignals-2 \cite{Bechtle:2020uwn},
for $\Delta \chi^2$ corresponding to $3 \sigma$ and $2 \sigma$, respectively.
Notice that there is a quantitative but not a qualitative difference between
the two different methods of including the signal strengths.
We see in Fig.~\ref{fig:1} that we can obtain
large pseudoscalar couplings, although maximal values are
excluded. For the wrong sign region we do obtain a valid region,
although we were not able to get points with $|c^o_b|> 0.001$.
As
explained in Ref.\cite{Biekotter:2024ykp}, the cancellations needed to
obey the present eEDM limit
of $4.1 \times 10^{-30}\, \textrm{e.cm}$ reported by
the JILA collaboration\cite{Roussy:2022cmp},
are quite large and depend
strongly on the precise value of the constants and at which scale they
are taken. In Ref.\cite{Biekotter:2024ykp} two cases were studied:
one case where the masses were taken as pole masses;
and another case where everything was considered at the $M_Z$ scale.
This had a strong impact; for instance,
the possibility of large pseudoscalar couplings for the type-II C2HDM
disappeared when taking the constants at the $M_Z$ scale.
The implication is that the scale dependence and, thus,
the assumed theoretical errors in estimating the eEDM warrants a full
independent and thorough analysis.
This lies well beyond the scope of the current work.

Also included is the latest data of direct searches for CP-violation
using angular 
correlations in decay planes of $\tau$ leptons produced in Higgs boson
decays $h_{125}\to
\tau\bar{\tau}$\cite{CMS:2021sdq,ATLAS:2022akr}. The best result is
from ATLAS\cite{ATLAS:2022akr},
setting an upper limit of $ \alpha_{h\tau\tau} < 34^\circ $ on the
effective mixing angle between the CP-even 
and CP-odd Yukawa coupling. This excludes the gray regions on the
plots. 

\begin{figure}[h!]
  \centering
  \begin{tabular}{cc}
    \includegraphics[width=0.48\textwidth]{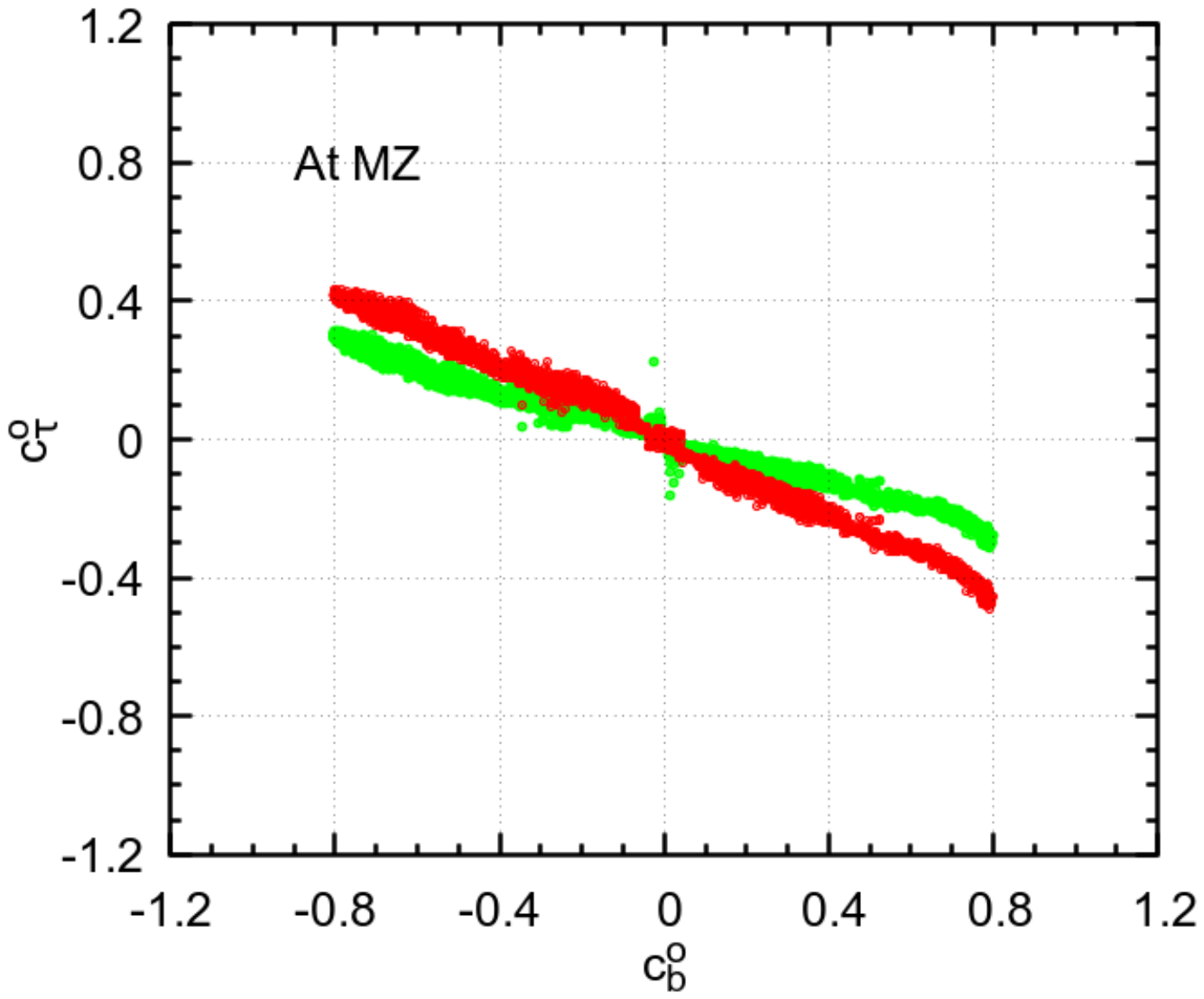}
  \end{tabular}
  \caption{We plot $c^o_\tau$ vs. $c^0_b$  for  masses at the $M_Z$ scale.
In red we plot all the points that we have generated and displayed
in Fig.~\ref{fig:1} and Fig.~\ref{fig:1b}, and in green the value
of $c^o_\tau$ obtained from Eq.~\eqref{anticorr}. }
  \label{fig:corr}
\end{figure}

One feature of Fig.~\ref{fig:1}, is that there are small regions
excluded by the cuts on $ \alpha_{h\tau\tau}$ for the case of the
parameters chosen at the scale $M_Z$. There are two points
that deserve a more detailed explanation. First, why the cuts on
$(c^e_\tau,c^o_\tau)$ have an impact on $(c^e_b,c^o_b)$, as we argued
before that for type-Z these would seem independent.
In fact such independence is not complete;
there are constraints to be satisfied,
as has been explained in section~\ref{s:HiggsCouplings}.
Indeed, Eq.~\eqref{anticorr} indicates that there is an anti-correlation
between $c^o_\tau$ and $c^o_b$ controlled by $\beta_1$. 
This is shown in Fig.~\ref{fig:corr}.

\begin{figure}[h!]
  \centering
  \begin{tabular}{cc}
    \includegraphics[width=0.48\textwidth]{aD-bD-AtMZ.pdf}
    &
    \includegraphics[width=0.48\textwidth]{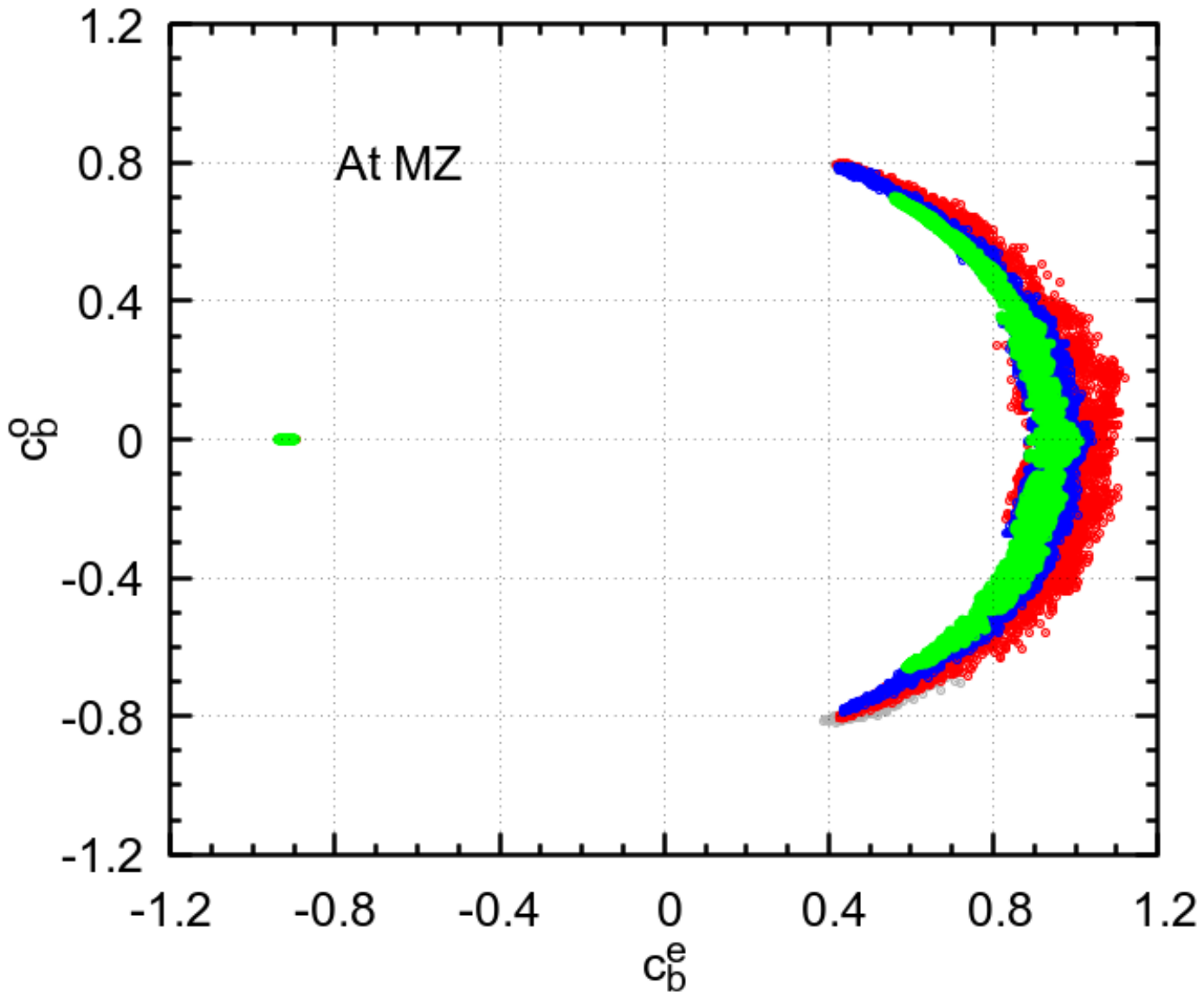}
  \end{tabular}
  \caption{In this plot the constants were taken
    at the $M_Z$ scale. On the left panel we show for comparison the
    right panel of Fig.~\ref{fig:1}, while in the right panel we show
    the result starting from a different initial point. See text for
    more details.}
  \label{fig:1b}
\end{figure}

The second point of note is why the gray regions on the right panel of Fig.~\ref{fig:1} are not
symmetric. This is a very interesting point and it elucidates what we
have said before about the covering of the parameter space with our
method of starting from the real 3HDM points. In Fig.~\ref{fig:1b} on
the left panel we
show, for comparison, again the right panel of Fig.~\ref{fig:1}. On
the right panel we show the result where we have started from a
different set of real 3HDM points. We see that now the regions are
almost symmetric. This example illustrates that all the points that we
show are good C3HDM points, but certainly there are other good C3HDM points
that we can not access in our method.
A final comment is in order. When we make a two dimensional plot, in
these models with a large number of parameters, we are superimposing
points that come from distinct regions in parameter space. If we look
for another characteristic (in our case the value of $
\alpha_{h\tau\tau}$) some will obey the constraint, others not. So we
have to be careful while drawing conclusions from these plots.

\begin{figure}[h!]
  \centering
  \begin{tabular}{cc}
    \includegraphics[width=0.48\textwidth]{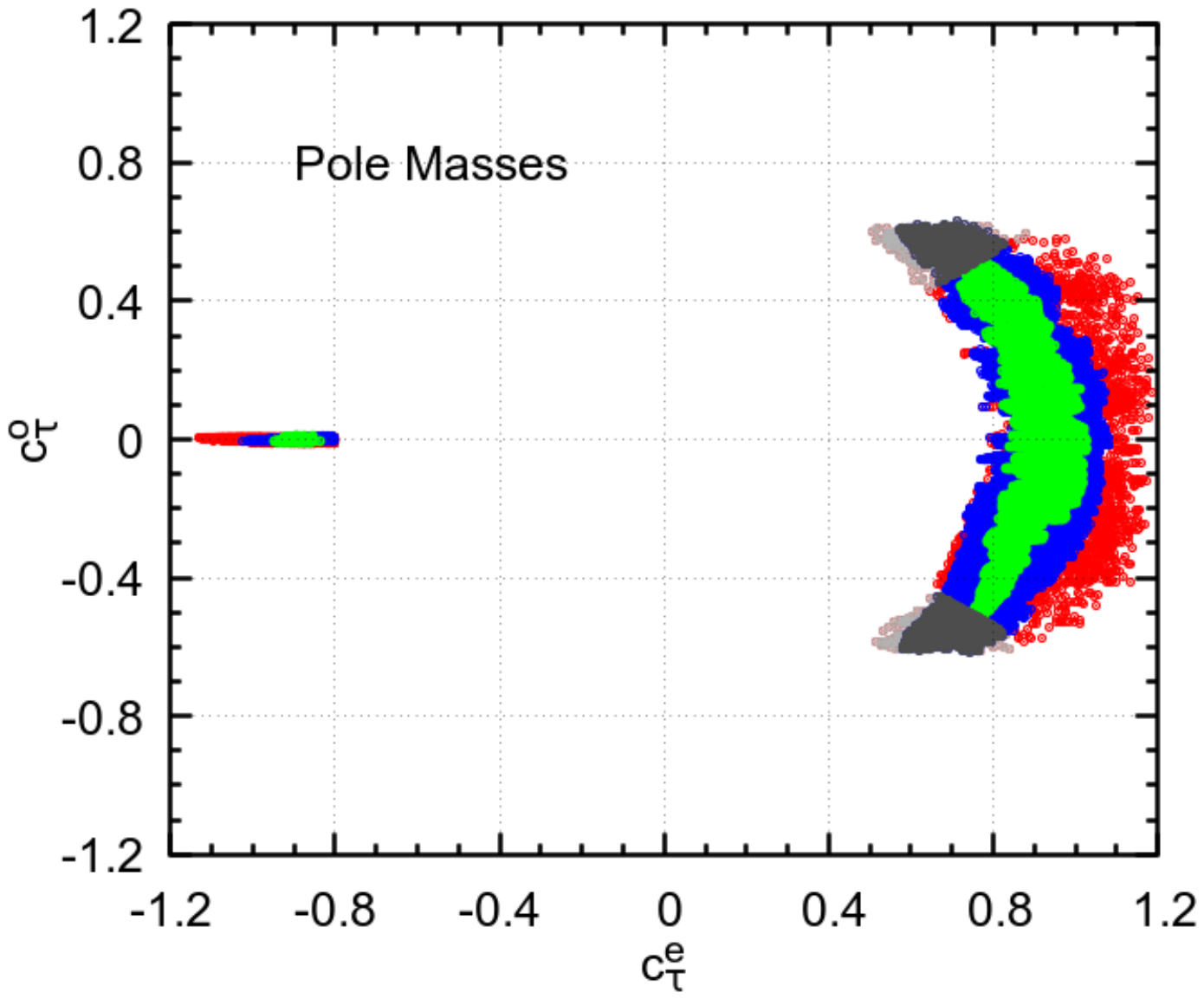}
    &
    \includegraphics[width=0.48\textwidth]{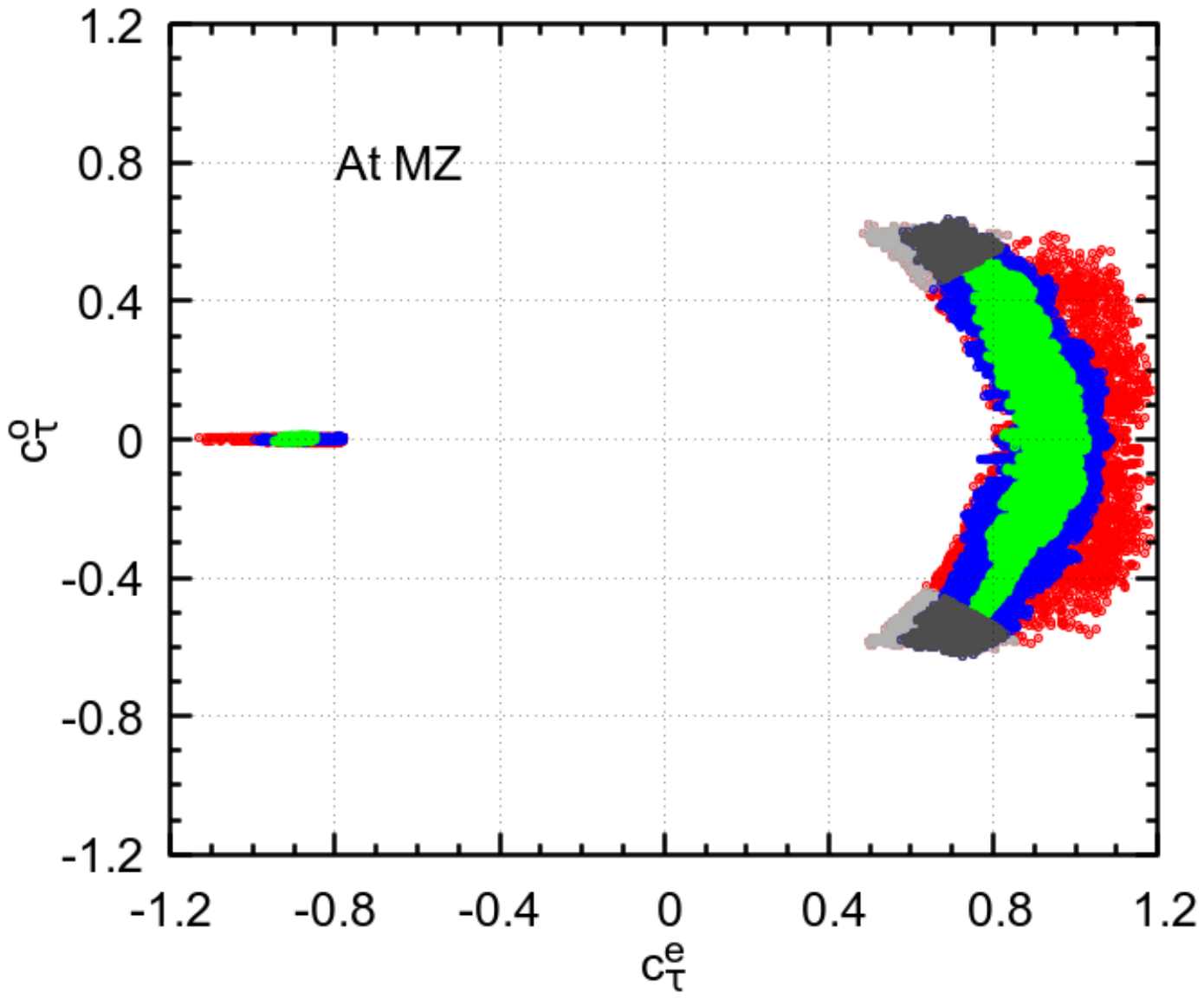}
  \end{tabular}
  \caption{$c^o_\tau$ vs. $c^e_\tau$. The constraints on $h_{125}\to
    \tau\bar{\tau}$ are included as the gray regions. See text for color code and more details.}
  \label{fig:2}
\end{figure}

In Fig.~\ref{fig:2} we show our results for the scalar and
pseudoscalar coupling of the charged leptons. The conclusions are
similar, although here we can not obtain as large a pseudoscalar
component as for the down-type quarks.

\begin{figure}[h!]
  \centering
  \begin{tabular}{cc}
    \includegraphics[width=0.56\textwidth]{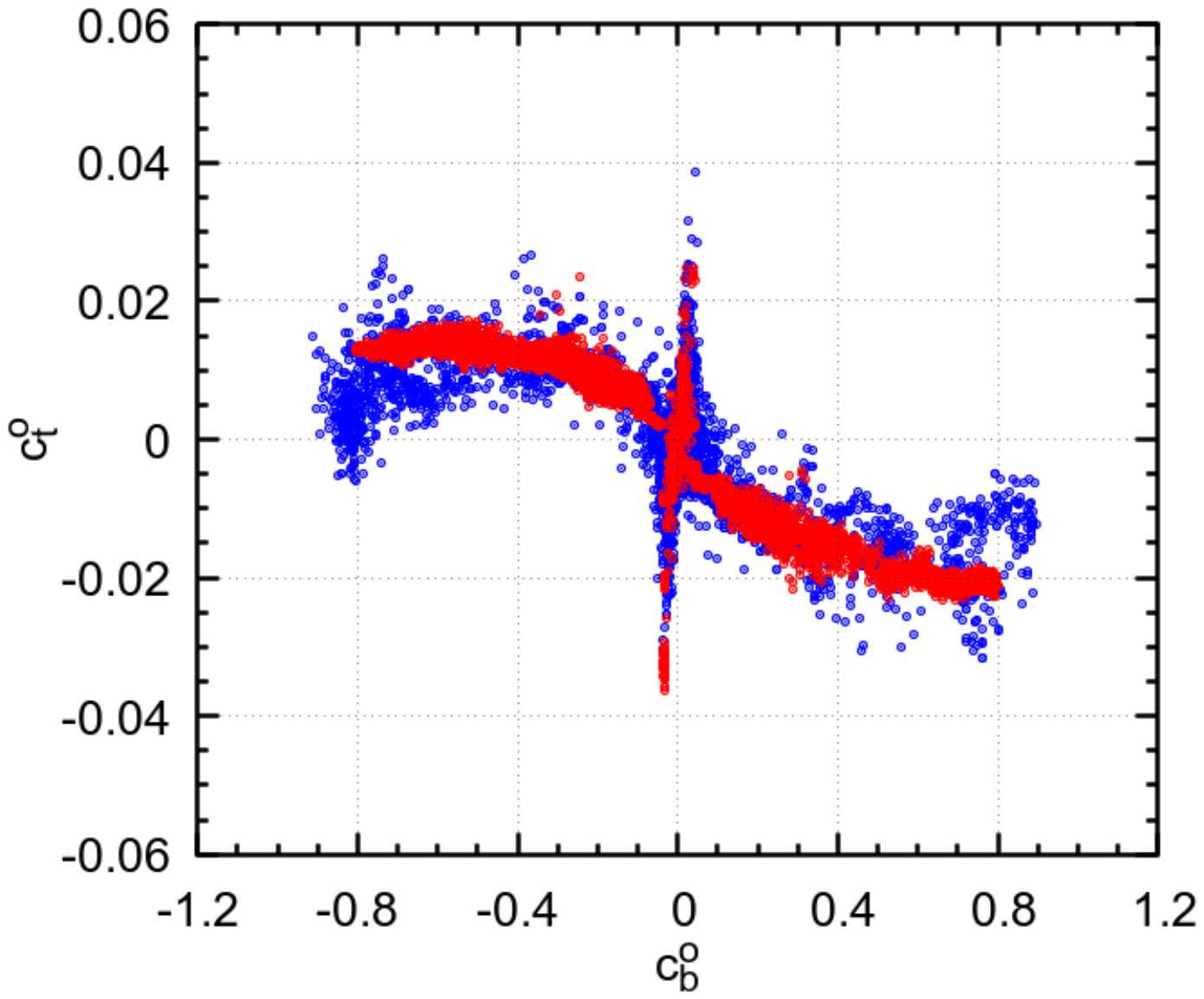}
  \end{tabular}
  \caption{$c^o_t$ vs. $c^o_b$. The points in blue (red) are obtained when the
eEDM is calculated with pole masses (masses at $M_Z$ scale). }
  \label{fig:ctocbo}
\end{figure}

It is also interesting to consider the correlation between
the pseudoscalar component for the down-type quarks and
the pseudoscalar component for the up-type quarks, shown in
Fig.~\ref{fig:ctocbo}.
As before, we consider two cases in the eEDM calculation:
one case where the masses were taken as pole masses (in blue);
and another case where everything was considered at the $M_Z$ scale (in red).
The situation is different in the two cases. When the quantities involved in the
eEDM calculation are taken at $M_Z$, it is only possible to attain a 
$c_b^o$ of order $0.8$, if $- c_t^o \sim c_b^o/40 \neq 0$. In contrast,
when the eEDM is calculated with pole masses, one can obtain 
$c_b^o \sim 0.8$, even if $c_t^o \sim 0$.
Thus, within the various methodology choices, there is room to have 
a coupling of $h_{125}$ to the top be exclusively scalar
while its coupling to the bottom quark is mostly pseudoscalar.


\section{Conclusions}
\label{sec:conclusions}

 The possibility that all fermions of a given charge couple to a different scalar doublet, the so-called type-Z models, occurs for
multi Higgs models with $N \geq 3$, the simplest example being the
3HDM. This has been extensively studied in the literature for the case
when all the parameters of the scalar potential are real.
In this work we consider the case with a generic vev where (some) parameters in the potential
are complex: the C3HDM. This has never been studied before, and
presents the possibility of addressing two important issues. 
Firstly, due to the fact that the Yukawa Lagrangian is already complex
because of the (complex) CKM matrix, the renormalization will require that the scalar
potential is also complex\cite{Fontes:2021znm}.
Secondly, the C3HDM can address the possibility of having large pseudoscalar
couplings for the fermions.
Recently, this possibility has been almost
ruled out in the C2HDM\cite{Biekotter:2024ykp},
due to the recent data on the eEDM and the searches at LHC.
One important constraint came from the measurement
of the decay $h_{125}\to \tau\bar{\tau}$.
In the C3HDM, as each right-handed fermion of a given charge
couples with only one Higgs doublet, we have more freedom.

We have introduced a parameterization for the rotation matrices needed
for the diagonalization of the neutral and charged scalars. This is
not trivial, as now all the five neutral scalars lack definite CP
properties and mix. 
This new formalism holds for Higgs-fermion couplings in any generic NFC 3HDM.
We then investigate the current bounds on the
C3HDM with $Z_2\times Z_2$ symmetry. We perform an up-to-date analysis,
including the latest data for the 125GeV Higgs \cite{ATLAS:2022vkf,CMS:2022dwd},
bounds on new scalars through the \texttt{HiggsTools} code
\cite{Bahl:2022igd}, and the very important theoretical
constraints.

We use the theoretical bounds from unitarity \cite{Bento:2022vsb} and
BFB \cite{Boto:2022uwv} and the standard procedure\cite{Grimus:2007if}
for the $S,T,U$
precison observables.  We stress the
importance of using the most recent LHC bounds, which constrain
severely the allowed parameter space. Also the constraints from the
latest bound on the eEDM\cite{Roussy:2022cmp} play an important
role. Like in the C2HDM\cite{Biekotter:2024ykp}, large cancellations
are needed which in turn display a dependence on the precise values taken for the
physical constants.

All these constraints make a random scan of the parameter space not
viable. So we followed the strategy of importing good points from the
real 3HDM and evolved from there, enlarging the pseudoscalar components
as much as possible. We have shown that, although the maximal CP-odd case
($c^e_b=0,c^o_b=\pm 1$) is excluded, a value close to
$c^e_b=0.4$ and $c^o_b=\pm 0.8$, is still allowed. For the case of the
$\tau$ lepton, the measurement of the CP properties of the $\tau$
lepton at the LHC\cite{CMS:2021sdq,ATLAS:2022akr} imposes stronger
constraints. 
Thus, the prospect of the CP-even/CP-odd $t$/$b$ possibility deserves continued
experimental exploration.

In the current procedure, one cannot be absolutely certain of
covering all regions of the parameter space away from the real
3HDM.
In the future,
we intend to apply the techniques developed in
Refs.\cite{deSouza:2022uhk,Romao:2024gjx} to explore fully the C3HDM
parameter space.

\section*{Acknowledgments}
\noindent
	This work is supported in part by the Portuguese Funda\c{c}\~{a}o
para a Ci\^{e}ncia e Tecnologia\/ (FCT) under Contracts
CERN/FIS-PAR/0002/2021, CERN/FIS-PAR/0008/2019, UIDB/00777/2020, and UIDP/00777/2020\,;
these projects are partially funded through POCTI (FEDER),
	COMPETE, QREN, and the EU. The work of R. Boto is also supported
	by FCT with the PhD grant PRT/BD/152268/2021.

\bibliographystyle{JHEP}
\bibliography{BFB}

\newpage
\appendix

\section{Some auxiliar matrices}\label{app:parametrization}

The z-tensor of quartic parameter is
\footnote{In this representation of $z_{ijkl}$,
one of the pairs of indices, $ij$/$kl$, refers to the entries of
the larger $3\times 3$ matrix,
while the other refers to the smaller $3\times 3$ matrices;
the order is not relevant due to the symmetry $z_{ijkl}=z_{klij}$.}
\begin{equation}\label{e:ztensor}
	z = 
	\begin{pmatrix}
		\begin{pmatrix}
			\lambda_1 & 0 & 0 \\
			0 & \lambda_4/2 & 0 \\
			0 & 0 & \lambda_5/2 \\
		\end{pmatrix}
		&
		\begin{pmatrix}
			0 & \lambda_{10} & 0 \\
			\lambda_7/2 & 0 & 0 \\
			0 & 0 & 0 \\
		\end{pmatrix}
		&
		\begin{pmatrix}
			0 & 0 & \lambda_{11} \\
			0 & 0 & 0 \\
			\lambda_8/2 & 0 & 0 \\
		\end{pmatrix}
		\\
		\begin{pmatrix}
			0 & \lambda_7/2 & 0 \\
			\lambda_{10}^* & 0 & 0 \\
			0 & 0 & 0 \\
		\end{pmatrix}
		&
		\begin{pmatrix}
			\lambda_4/2 & 0 & 0 \\
			0 & \lambda_2 & 0 \\
			0 & 0 & \lambda_6/2 \\
		\end{pmatrix}
		&
		\begin{pmatrix}
			0 & 0 & 0 \\
			0 & 0 & \lambda_{12} \\
			0 & \lambda_9/2 & 0 \\
		\end{pmatrix}
		\\
		\begin{pmatrix}
			0 & 0 & \lambda_8/2 \\
			0 & 0 & 0 \\
			\lambda_{11}^* & 0 & 0 \\
		\end{pmatrix}
		&
		\begin{pmatrix}
			0 & 0 & 0 \\
			0 & 0 & \lambda_9/2 \\
			0 & \lambda_{12}^* & 0 \\
		\end{pmatrix}
		&
		\begin{pmatrix}
			\lambda_5/2 & 0 & 0 \\
			0 & \lambda_6/2 & 0 \\
			0 & 0 & \lambda_3 \\
		\end{pmatrix}
	\end{pmatrix}
	\;,
\end{equation}
giving us the matrices 
from which we can obtain the matrices $A$, $B$, $C$ for our model
\begin{equation}\label{e:matA}
	A = \boldsymbol{\mu}  + 
	\left(\begin{matrix}\lambda_{1} v_{1}^{2} + \frac{\lambda_{4} v_{2}^{2}}{2} + \frac{\lambda_{5} v_{3}^{2}}{2} & \frac{\lambda_{7} v_{1} v_{2}}{2} + \lambda_{10} v_{1} v_{2} & \frac{\lambda_{8} v_{1} v_{3}}{2} + \lambda_{11} v_{1} v_{3}\\\frac{\lambda_{7} v_{1} v_{2}}{2} + v_{1} v_{2} \lambda_{10}^* & \lambda_{2} v_{2}^{2} + \frac{\lambda_{4} v_{1}^{2}}{2} + \frac{\lambda_{6} v_{3}^{2}}{2} & \frac{\lambda_{9} v_{2} v_{3}}{2} + \lambda_{12} v_{2} v_{3}\\\frac{\lambda_{8} v_{1} v_{3}}{2} + v_{1} v_{3} \lambda_{11}^* & \frac{\lambda_{9} v_{2} v_{3}}{2} + v_{2} v_{3} \lambda_{12}^* & \lambda_{3} v_{3}^{2} + \frac{\lambda_{5} v_{1}^{2}}{2} + \frac{\lambda_{6} v_{2}^{2}}{2}\end{matrix}\right) = M^2_{ch}
	\;,
\end{equation}
\begin{equation}\label{e:matB}
	B =
	\left(\begin{matrix}\lambda_{1} v_{1}^{2} + \frac{\lambda_{7} v_{2}^{2}}{2} + \frac{\lambda_{8} v_{3}^{2}}{2} & \frac{\lambda_{4} v_{1} v_{2}}{2} + \lambda_{10} v_{1} v_{2} & \frac{\lambda_{5} v_{1} v_{3}}{2} + \lambda_{11} v_{1} v_{3}\\\frac{\lambda_{4} v_{1} v_{2}}{2} + v_{1} v_{2} {\lambda_{10}^*} & \lambda_{2} v_{2}^{2} + \frac{\lambda_{7} v_{1}^{2}}{2} + \frac{\lambda_{9} v_{3}^{2}}{2} & \frac{\lambda_{6} v_{2} v_{3}}{2} + \lambda_{12} v_{2} v_{3}\\\frac{\lambda_{5} v_{1} v_{3}}{2} + v_{1} v_{3} {\lambda_{11}^*} & \frac{\lambda_{6} v_{2} v_{3}}{2} + v_{2} v_{3} {\lambda_{12}^*} & \lambda_{3} v_{3}^{2} + \frac{\lambda_{8} v_{1}^{2}}{2} + \frac{\lambda_{9} v_{2}^{2}}{2}\end{matrix}\right)
	\;,
\end{equation}
\begin{equation}\label{e:matC}
	C =
	\left(\begin{matrix}\lambda_{1} v_{1}^{2} + v_{2}^{2} {\lambda_{10}^*} + v_{3}^{2} {\lambda_{11}^*} & \frac{\lambda_{4} v_{1} v_{2}}{2} + \frac{\lambda_{7} v_{1} v_{2}}{2} & \frac{\lambda_{5} v_{1} v_{3}}{2} + \frac{\lambda_{8} v_{1} v_{3}}{2}\\\frac{\lambda_{4} v_{1} v_{2}}{2} + \frac{\lambda_{7} v_{1} v_{2}}{2} & \lambda_{2} v_{2}^{2} + \lambda_{10} v_{1}^{2} + v_{3}^{2} {\lambda_{12}^*} & \frac{\lambda_{6} v_{2} v_{3}}{2} + \frac{\lambda_{9} v_{2} v_{3}}{2}\\\frac{\lambda_{5} v_{1} v_{3}}{2} + \frac{\lambda_{8} v_{1} v_{3}}{2} & \frac{\lambda_{6} v_{2} v_{3}}{2} + \frac{\lambda_{9} v_{2} v_{3}}{2} & \lambda_{3} v_{3}^{2} + \lambda_{11} v_{1}^{2} + \lambda_{12} v_{2}^{2}\end{matrix}\right)
	\;,
\end{equation}
from which the mass matrices can be written.

\section{Mass matrices of the softly-broken \texorpdfstring{$Z_2\times Z_2$}{} symmetric 3HDM}\label{app:mass_matrices}

\begin{equation}
	\left(M^{2\prime}_{n}\right)_{1,1} = - \frac{\Re(\mu_{12}) s_{\beta_1}}{c_{\beta_1}} - \frac{\Re(\mu_{13}) s_{\beta_2}}{c_{\beta_1} c_{\beta_2}} + 2 \lambda_{1} c_{\beta_1}^{2} c_{\beta_2}^{2} v^{2}
\end{equation}

\begin{equation}
	\left(M^{2\prime}_{n}\right)_{1,2} = 2 \Re(\lambda_{10}) c_{\beta_1} c_{\beta_2}^{2} s_{\beta_1} v^{2} + \Re(\mu_{12}) + \lambda_{4} c_{\beta_1} c_{\beta_2}^{2} s_{\beta_1} v^{2} + \lambda_{7} c_{\beta_1} c_{\beta_2}^{2} s_{\beta_1} v^{2}
\end{equation}

\begin{equation}
	\left(M^{2\prime}_{n}\right)_{1,3} = 2 \Re(\lambda_{11}) c_{\beta_1} c_{\beta_2} s_{\beta_2} v^{2} + \Re(\mu_{13}) + \lambda_{5} c_{\beta_1} c_{\beta_2} s_{\beta_2} v^{2} + \lambda_{8} c_{\beta_1} c_{\beta_2} s_{\beta_2} v^{2}
\end{equation}

\begin{equation}
	\left(M^{2\prime}_{n}\right)_{1,4} = - \Im(\lambda_{10}) c_{\beta_2}^{2} s_{\beta_1} v^{2} \left(2 - s_{\beta_1}^{2}\right) - \Im(\lambda_{11}) s_{\beta_1} s_{\beta_2}^{2} v^{2} - \Im(\mu_{12}) c_{\beta_1}
\end{equation}

\begin{equation}
	\left(M^{2\prime}_{n}\right)_{1,5} = \frac{\Im(\lambda_{10}) c_{\beta_1} c_{\beta_2}^{2} s_{\beta_1}^{2} v^{2}}{s_{\beta_2}} - \Im(\lambda_{11}) c_{\beta_1} s_{\beta_2} v^{2} + \frac{\Im(\mu_{12}) s_{\beta_1}}{s_{\beta_2}}
\end{equation}

\begin{equation}
	\left(M^{2\prime}_{n}\right)_{2,2} = - \frac{\Re(\mu_{12}) c_{\beta_1}}{s_{\beta_1}} - \frac{\Re(\mu_{23}) s_{\beta_2}}{c_{\beta_2} s_{\beta_1}} + 2 \lambda_{2} c_{\beta_2}^{2} s_{\beta_1}^{2} v^{2}
\end{equation}

\begin{equation}
	\left(M^{2\prime}_{n}\right)_{2,3} = 2 \Re(\lambda_{12}) c_{\beta_2} s_{\beta_1} s_{\beta_2} v^{2} + \Re(\mu_{23}) + \lambda_{6} c_{\beta_2} s_{\beta_1} s_{\beta_2} v^{2} + \lambda_{9} c_{\beta_2} s_{\beta_1} s_{\beta_2} v^{2}
\end{equation}

\begin{equation}
	\left(M^{2\prime}_{n}\right)_{2,4} = \Im(\lambda_{10}) c_{\beta_1} c_{\beta_2}^{2} v^{2} \left(- s_{\beta_1}^{2} - 1\right) + \Im(\lambda_{12}) c_{\beta_1} s_{\beta_2}^{2} v^{2} - \Im(\mu_{12}) s_{\beta_1}
\end{equation}

\begin{equation}
	\left(M^{2\prime}_{n}\right)_{2,5} = - \frac{\Im(\lambda_{10}) c_{\beta_1}^{2} c_{\beta_2}^{2} s_{\beta_1} v^{2}}{s_{\beta_2}} - \Im(\lambda_{12}) s_{\beta_1} s_{\beta_2} v^{2} - \frac{\Im(\mu_{12}) c_{\beta_1}}{s_{\beta_2}}
\end{equation}

\begin{equation}
	\left(M^{2\prime}_{n}\right)_{3,3} = - \frac{\Re(\mu_{13}) c_{\beta_1} c_{\beta_2}}{s_{\beta_2}} - \frac{\Re(\mu_{23}) c_{\beta_2} s_{\beta_1}}{s_{\beta_2}} + 2 \lambda_{3} s_{\beta_2}^{2} v^{2}
\end{equation}

\begin{equation}\begin{split}
		\left(M^{2\prime}_{n}\right)_{3,4} =& \frac{\Im(\lambda_{10}) c_{\beta_1} c_{\beta_2}^{3} s_{\beta_1} v^{2}}{s_{\beta_2}} - \Im(\lambda_{11}) c_{\beta_1} c_{\beta_2} s_{\beta_1} s_{\beta_2} v^{2} \\& +\Im(\lambda_{12}) c_{\beta_1} c_{\beta_2} s_{\beta_1} s_{\beta_2} v^{2} + \frac{\Im(\mu_{12}) c_{\beta_2}}{s_{\beta_2}}
\end{split}\end{equation}

\begin{equation}
	\left(M^{2\prime}_{n}\right)_{3,5} = - \Im(\lambda_{11}) c_{\beta_1}^{2} c_{\beta_2} v^{2} - \Im(\lambda_{12}) c_{\beta_2} s_{\beta_1}^{2} v^{2}
\end{equation}

\begin{equation}\begin{split}
		\left(M^{2\prime}_{n}\right)_{4,4} =& - 2 \Re(\lambda_{10}) c_{\beta_2}^{2} v^{2} - 2 \Re(\lambda_{11}) s_{\beta_1}^{2} s_{\beta_2}^{2} v^{2} - 2 \Re(\lambda_{12}) c_{\beta_1}^{2} s_{\beta_2}^{2} v^{2}  \\ & -\frac{\Re(\mu_{12})}{c_{\beta_1} s_{\beta_1}} - \frac{\Re(\mu_{13}) s_{\beta_1}^{2} s_{\beta_2}}{c_{\beta_1} c_{\beta_2}} - \frac{\Re(\mu_{23}) c_{\beta_1}^{2} s_{\beta_2}}{c_{\beta_2} s_{\beta_1}}
\end{split}\end{equation}

\begin{equation}
	\left(M^{2\prime}_{n}\right)_{4,5} = - 2 \Re(\lambda_{11}) c_{\beta_1} s_{\beta_1} s_{\beta_2} v^{2} + 2 \Re(\lambda_{12}) c_{\beta_1} s_{\beta_1} s_{\beta_2} v^{2} - \frac{\Re(\mu_{13}) s_{\beta_1}}{c_{\beta_2}} + \frac{\Re(\mu_{23}) c_{\beta_1}}{c_{\beta_2}}
\end{equation}

\begin{equation}
	\left(M^{2\prime}_{n}\right)_{5,5} = - 2 \Re(\lambda_{11}) c_{\beta_1}^{2} v^{2} - 2 \Re(\lambda_{12}) s_{\beta_1}^{2} v^{2} - \frac{\Re(\mu_{13}) c_{\beta_1}}{c_{\beta_2} s_{\beta_2}} - \frac{\Re(\mu_{23}) s_{\beta_1}}{c_{\beta_2} s_{\beta_2}}
\end{equation}

\begin{equation}\begin{split}
		\left(M^{2\prime}_{ch}\right)_{1,1} =& - \Re(\lambda_{10}) c_{\beta_2}^{2} v^{2} - \Re(\lambda_{11}) s_{\beta_1}^{2} s_{\beta_2}^{2} v^{2} - \Re(\lambda_{12}) c_{\beta_1}^{2} s_{\beta_2}^{2} v^{2} - \frac{\Re(\mu_{12})}{c_{\beta_1} s_{\beta_1}} \\& - \frac{\Re(\mu_{13}) s_{\beta_1}^{2} s_{\beta_2}}{c_{\beta_1} c_{\beta_2}}  - \frac{\Re(\mu_{23}) c_{\beta_1}^{2} s_{\beta_2}}{c_{\beta_2} s_{\beta_1}} - \frac{\lambda_{7} c_{\beta_2}^{2} v^{2}}{2} - \frac{\lambda_{8} s_{\beta_1}^{2} s_{\beta_2}^{2} v^{2}}{2} - \frac{\lambda_{9} c_{\beta_1}^{2} s_{\beta_2}^{2} v^{2}}{2}
\end{split}\end{equation}

\begin{equation}
	\left(M^{2\prime}_{ch}\right)_{2,2} = - \Re(\lambda_{11}) c_{\beta_1}^{2} v^{2} - \Re(\lambda_{12}) s_{\beta_1}^{2} v^{2} - \frac{\Re(\mu_{13}) c_{\beta_1}}{c_{\beta_2} s_{\beta_2}} - \frac{\Re(\mu_{23}) s_{\beta_1}}{c_{\beta_2} s_{\beta_2}} - \frac{\lambda_{8} c_{\beta_1}^{2} v^{2}}{2} - \frac{\lambda_{9} s_{\beta_1}^{2} v^{2}}{2}
\end{equation}

\begin{equation}\begin{split}
		\Re{{\left(M^{{2\prime}}_{{ch}}\right)_{{1,2}}}} =& - \Re(\lambda_{11}) c_{\beta_1} s_{\beta_1} s_{\beta_2} v^{2} + \Re(\lambda_{12}) c_{\beta_1} s_{\beta_1} s_{\beta_2} v^{2} - \frac{\Re(\mu_{13}) s_{\beta_1}}{c_{\beta_2}} \\&  + \frac{\Re(\mu_{23}) c_{\beta_1}}{c_{\beta_2}} - \frac{\lambda_{8} c_{\beta_1} s_{\beta_1} s_{\beta_2} v^{2}}{2} + \frac{\lambda_{9} c_{\beta_1} s_{\beta_1} s_{\beta_2} v^{2}}{2}
\end{split}\end{equation}

\begin{equation}
	\Im{{\left(M^{{2\prime}}_{{ch}}\right)_{{1,2}}}} = \frac{\Im(\lambda_{10}) c_{\beta_1} c_{\beta_2}^{2} s_{\beta_1} v^{2}}{s_{\beta_2}} + \frac{\Im(\mu_{12})}{s_{\beta_2}}
\end{equation}


\section{Rotation to Mass Basis}\label{a:rotation_matrices}

In order to analyze the couplings of the scalar particles, and some of the theoretical constraints on the parameter space, we need to write the matrix that rotates the original (symmetry) basis fields to the mass basis fields. 
The charged would-be Goldstone boson $G^+ = w_1^{+\prime}$ is automatically the first charged scalar, and it is useful to also put the neutral would-be Goldstone boson $G^0=z_1^\prime$ in the first position, so we write
\begin{equation}
	\begin{pmatrix}
		G^+ \\ H_1^+ \\ H_2^+
	\end{pmatrix}
	=
	\begin{pmatrix}
     1 & 0 & 0  \\
    0 &\multicolumn{2}{c}{\multirow{2}{*}{ $ W$}}\\ 
    0  &
    \end{pmatrix}
	\begin{pmatrix}
		G^+ \\ w_2^{+\prime} \\ w_3^{+\prime}
	\end{pmatrix}
	= 
	\begin{pmatrix}
     1 & 0 & 0  \\
    0 &\multicolumn{2}{c}{\multirow{2}{*}{ $ W$}}\\ 
    0  &
    \end{pmatrix}
	\;R_H\;
	\begin{pmatrix}
		w_1^+ \\ w_2^+ \\ w_3^+
	\end{pmatrix}
	=
	U^\dag
	\begin{pmatrix}
		w_1^+ \\ w_2^+ \\ w_3^+
	\end{pmatrix}
	\;,
\end{equation}
where we define the matrix $U$ as

\begin{equation}\label{matrixU}
	\begin{split}
		U= R_H^\dag 
		\begin{pmatrix}
     1 & 0 & 0  \\
    0 &\multicolumn{2}{c}{\multirow{2}{*}{ $ W^\dag$}}\\ 
    0  &
    \end{pmatrix}
		=
		\begin{pmatrix}
			c_{\beta_2} c_{\beta_1} & -s_{\beta_1} & -s_{\beta_2} c_{\beta_1} \\
			c_{\beta_2} s_{\beta_1} & c_{\beta_1} & -s_{\beta_2} s_{\beta_1} \\
			s_{\beta_2} & 0 & c_{\beta_2} \\
		\end{pmatrix}
		\begin{pmatrix}
			1 & 0 & 0  \\
			0 & c_\theta e^{-i \varphi} & -s_\theta e^{-i \varphi}  \\
			0 & s_\theta e^{i \varphi} & c_\theta e^{i \varphi}  
		\end{pmatrix}
		\\= 
		\begin{pmatrix}
			c_{\beta_2} c_{\beta_1} & -s_{\beta_1}c_\theta e^{-i \varphi} -s_{\beta_2} c_{\beta_1}s_\theta e^{i \varphi} & s_{\beta_1}s_\theta e^{-i \varphi} -s_{\beta_2} c_{\beta_1}c_\theta e^{i \varphi} \\
			c_{\beta_2} s_{\beta_1} & c_{\beta_1}c_\theta e^{-i \varphi} -s_{\beta_2} s_{\beta_1}s_\theta e^{i \varphi} & -c_{\beta_1}s_\theta e^{-i \varphi} -s_{\beta_2} s_{\beta_1}c_\theta e^{i \varphi} \\
			s_{\beta_2} & c_{\beta_2} s_\theta e^{i \varphi} & c_{\beta_2} c_\theta e^{i \varphi} 
		\end{pmatrix}
		\;,
	\end{split}
\end{equation}
and
\begin{equation}\label{matrixQ}
	\begin{split}
		\begin{pmatrix}
			G^0 \\ h_1 \\ h_2 \\ h_3 \\ h_4 \\ h_5
		\end{pmatrix}
		&=
	\begin{pmatrix}
     1 & 0 & 0 & 0 & 0& 0 \\
    0 &\multicolumn{5}{c}{\multirow{5}{*}{ $R$}}\\ 
    0 &  &  &  &  &   \\ 
    0 &  &  &  &  &   \\ 
    0 &  &  &  &  &   \\ 
    0 &  &  &  &  &   \\ 
    \end{pmatrix}
		\begin{pmatrix}
			G^0 \\ x_1 \\ x_2 \\ x_3 \\ z_2^\prime \\ z_3^\prime
		\end{pmatrix}
		=
		\begin{pmatrix}
     1 & 0 & 0 & 0 & 0& 0 \\
    0 &\multicolumn{5}{c}{\multirow{5}{*}{ $R$}}\\ 
    0 &  &  &  &  &   \\ 
    0 &  &  &  &  &   \\ 
    0 &  &  &  &  &   \\ 
    0 &  &  &  &  &   \\ 
    \end{pmatrix}
		\begin{pmatrix}
			0 & 0 & 0 & 1 & 0 & 0\\
			1 & 0 & 0 & 0 & 0 & 0\\
			0 & 1 & 0 & 0 & 0 & 0\\
			0 & 0 & 1 & 0 & 0 & 0\\
			0 & 0 & 0 & 0 & 1 & 0\\
			0 & 0 & 0 & 0 & 0 & 1\\
		\end{pmatrix}
		\begin{pmatrix}
			x_1 \\ x_2 \\ x_3 \\ G^0 \\ z_2^\prime \\ z_3^\prime
		\end{pmatrix}
		\\
		&=
		\begin{pmatrix}
     1 & 0 & 0 & 0 & 0& 0 \\
    0 &\multicolumn{5}{c}{\multirow{5}{*}{ $R$}}\\ 
    0 &  &  &  &  &   \\ 
    0 &  &  &  &  &   \\ 
    0 &  &  &  &  &   \\ 
    0 &  &  &  &  &   \\ 
    \end{pmatrix}
		\begin{pmatrix}
			0 & 0 & 0 & 1 & 0 & 0\\
			1 & 0 & 0 & 0 & 0 & 0\\
			0 & 1 & 0 & 0 & 0 & 0\\
			0 & 0 & 1 & 0 & 0 & 0\\
			0 & 0 & 0 & 0 & 1 & 0\\
			0 & 0 & 0 & 0 & 0 & 1\\
		\end{pmatrix}
		\begin{pmatrix}
     \multicolumn{3}{c}{\multirow{3}{*}{ $\one$}} &  \multicolumn{3}{c}{\multirow{3}{*}{ $0$}} \\    
     &  &  &  &  &   \\ 
         &  &  &  &  &   \\ 
 \multicolumn{3}{c}{\multirow{3}{*}{ $0$}} &  \multicolumn{3}{c}{\multirow{3}{*}{ $R_H$}}  \\    
     &  &  &  &  &   \\ 
         &  &  &  &  &   \\ 
    \end{pmatrix}
		\begin{pmatrix}
			x_1 \\ x_2 \\ x_3 \\ z_1 \\ z_2 \\ z_3
		\end{pmatrix}
		\\
		&=
		\begin{pmatrix}
			0      &      0 &      0 & 1 &      0 &      0 \\
			R_{11} & R_{12} & R_{13} & 0 & R_{14} & R_{15} \\
			R_{21} & R_{22} & R_{23} & 0 & R_{24} & R_{25} \\
			R_{31} & R_{32} & R_{33} & 0 & R_{34} & R_{35} \\
			R_{41} & R_{42} & R_{43} & 0 & R_{44} & R_{45} \\
			R_{51} & R_{52} & R_{53} & 0 & R_{54} & R_{55} \\
		\end{pmatrix}
\begin{pmatrix}
     \multicolumn{3}{c}{\multirow{3}{*}{ $\one$}} &  \multicolumn{3}{c}{\multirow{3}{*}{ $0$}} \\    
     &  &  &  &  &   \\ 
         &  &  &  &  &   \\ 
 \multicolumn{3}{c}{\multirow{3}{*}{ $0$}}   &  c_{\beta_2} c_{\beta_1} & c_{\beta_2} s_{\beta_1} & s_{\beta_2} \\
			&  &  & -s_{\beta_1} & c_{\beta_1} & 0 \\
			&  &  & -s_{\beta_2} c_{\beta_1} & -s_{\beta_2} s_{\beta_1} & c_{\beta_2} \\
    \end{pmatrix}
		\begin{pmatrix}
			x_1 \\ x_2 \\ x_3 \\ z_1 \\ z_2 \\ z_3
		\end{pmatrix}
		\\
		&=
		\left(\begin{matrix}0 & 0 & 0 & c_{\beta_1} c_{\beta_2} & c_{\beta_2} s_{\beta_1} & s_{\beta_2}\\R_{11} & R_{12} & R_{13} & - R_{14} s_{\beta_1} - R_{15} c_{\beta_1} s_{\beta_2} & R_{14} c_{\beta_1} - R_{15} s_{\beta_1} s_{\beta_2} & R_{15} c_{\beta_2}\\R_{21} & R_{22} & R_{23} & - R_{24} s_{\beta_1} - R_{25} c_{\beta_1} s_{\beta_2} & R_{24} c_{\beta_1} - R_{25} s_{\beta_1} s_{\beta_2} & R_{25} c_{\beta_2}\\R_{31} & R_{32} & R_{33} & - R_{34} s_{\beta_1} - R_{35} c_{\beta_1} s_{\beta_2} & R_{34} c_{\beta_1} - R_{35} s_{\beta_1} s_{\beta_2} & R_{35} c_{\beta_2}\\R_{41} & R_{42} & R_{43} & - R_{44} s_{\beta_1} - R_{45} c_{\beta_1} s_{\beta_2} & R_{44} c_{\beta_1} - R_{45} s_{\beta_1} s_{\beta_2} & R_{45} c_{\beta_2}\\R_{51} & R_{52} & R_{53} & - R_{54} s_{\beta_1} - R_{55} c_{\beta_1} s_{\beta_2} & R_{54} c_{\beta_1} - R_{55} s_{\beta_1} s_{\beta_2} & R_{55} c_{\beta_2}\end{matrix}\right)
		\begin{pmatrix}
			x_1 \\ x_2 \\ x_3 \\ z_1 \\ z_2 \\ z_3
		\end{pmatrix}
\\
		&=
		Q
		\begin{pmatrix}
			x_1 \\ x_2 \\ x_3 \\ z_1 \\ z_2 \\ z_3
		\end{pmatrix}
		\;,
	\end{split}
\end{equation}
where we defined the matrix $Q$. It is also be useful to define the following matrix $V$ and write this as
\small
\begin{eqnarray}\label{matrixV}
	\begin{pmatrix}
		x_1 + i z_1 \\ x_2 + i z_2 \\ x_3 + i z_3
	\end{pmatrix}
	&=&
	\left(\begin{matrix}Q^{T}_{11} + i Q^{T}_{41} & Q^{T}_{12} + i Q^{T}_{42}  & Q^{T}_{13} + i Q^{T}_{43}  & Q^{T}_{14} + i Q^{T}_{44}  & Q^{T}_{15} + i Q^{T}_{45}  & Q^{T}_{16} + i Q^{T}_{46} \\Q^{T}_{21} + i Q
		^{T}_{51}  & Q^{T}_{22} + i Q^{T}_{52}  & Q^{T}_{23} + i Q^{T}_{53}  & Q^{T}_{24} + i Q^{T}_{54}  & Q^{T}_{25} + i Q^{T}_{55}  & Q^{T}_{26} + i Q^{T}_{56} \\Q^{T}_{31} + i Q^{T}_{61}  & Q^{T}_{32} + i  Q^{T}_{
			62}  & Q^{T}_{33} + i Q^{T}_{63}  & Q^{T}_{34} + i Q^{T}_{64}  & Q^{T}_{35} + i Q^{T}_{65}  & Q^{T}_{36} + i Q^{T}_{66} \end{matrix}\right)
	\begin{pmatrix}
		G^0 \\ h_1 \\ h_2 \\ h_3 \\ h_4 \\ h_5
	\end{pmatrix}
\nonumber\\
	&=&
	V
	\begin{pmatrix}
		G^0 \\ h_1 \\ h_2 \\ h_3 \\ h_4 \\ h_5
	\end{pmatrix}
	\;.
\end{eqnarray}

\end{document}